\documentclass[nonacm, acmsmall]{acmart}
\usepackage{array}
\usepackage{geometry}
\usepackage{enumitem}
\usepackage{tablefootnote}
\renewcommand\footnotetextcopyrightpermission[1]{} %
\AtBeginDocument{%
  }

\begin{document}

\newcommand{\TOOL}{\textsc{FareShare}}
\newcommand{\LU}{\textsc{DU}}

\title[\TOOL{}]{\TOOL{}: A Tool for Labor Organizers to Estimate Lost Wages and Contest Arbitrary AI and Algorithmic Deactivations}

\author{Varun Nagaraj Rao}
\affiliation{%
  \institution{Center for Information Technology Policy, Princeton University}
  \country{USA}
  }
\email{varunrao@princeton.edu}
\author{Samantha Dalal}
\affiliation{%
  \institution{University of Colorado Boulder}
  \country{USA}
  }
\email{samantha.dalal@colorado.edu}
\author{Andrew Schwartz}
\email{andrew@cornflowerlabs.com}
\affiliation{%
  \institution{Cornflower Labs}
  \country{USA}
  }
\author{Amna Liaqat}
\email{al0910@princeton.edu}
\affiliation{%
  \institution{Center for Information Technology Policy, Princeton University}
  \country{USA}
  }
\author{Dana Calacci}
\email{dcalacci@princeton.edu}
\affiliation{%
  \institution{Penn State University}
  \country{USA}
  }
\author{Andrés Monroy-Hernández}
\email{andresmh@princeton.edu}
\affiliation{%
  \institution{Center for Information Technology Policy, Princeton University}
  \country{USA}
  }

\renewcommand{\shortauthors}{Nagaraj Rao et al.}

\begin{abstract}
What happens when a rideshare driver is suddenly locked out of the platform connecting them to riders, wages, and daily work? Deactivation—the abrupt removal of gig workers' platform access—typically occurs through arbitrary AI and algorithmic decisions with little explanation or recourse. This represents one of the most severe forms of algorithmic control and often devastates workers' financial stability. Recent U.S. state policies now mandate appeals processes and recovering compensation during the period of wrongful deactivation based on past earnings. Yet, labor organizers still lack effective tools to support these complex, error-prone workflows. We designed \TOOL{}, a computational tool automating lost wage estimation for deactivated drivers, through a 6 month partnership with the State of Washington's largest rideshare labor union. Over the following 3 months, our field deployment of \TOOL{} registered 178 account signups. We observed that the tool could reduce lost wage calculation time by over 95\%, eliminate manual data entry errors, and enable legal teams to generate arbitration-ready reports more efficiently. Beyond these gains, the deployment also surfaced important socio-technical challenges around trust, consent, and tool adoption in high-stakes labor contexts.
\end{abstract}

\maketitle

\section{Introduction}

Deactivation—the sudden termination of a gig worker's platform access—has emerged as a critical yet understudied mechanism of algorithmic control~\cite{lee2015working,rosenblatAlgorithmicLaborInformation2016b} in the gig economy and one of the most severe and destabilizing of AI and algorithmic harms for workers~\cite{wapo2021deactivated,nyt2025deactivated, schwartz2023deactivation}. Prior work has found that rideshare platforms may restrict drivers' accounts based on acceptance rates, ratings, or alleged policy violations~\cite{cameron2022making}. These studies demonstrate that unlike traditional employment termination, deactivation in gig work functions as algorithmic firing of independent contractors, who ostensibly have freedom to make their own decisions but in reality lack legal protections against unfair dismissal. The problem worsens as these deactivations remain largely arbitrary, with a study from the State of Washington revealing that platforms overturn 80\% of deactivation decisions upon appeal~\cite{schwartz2023deactivation}. Deactivation thus operates in a regulatory gray zone, leaving workers with minimal recourse.

Recent state-level policies in the U.S., including State of Washington's HB 2076 and Colorado's SB24-075, aim to address this gap. These laws mandate that platforms provide deactivation notices, formal appeals processes, and compensation for wrongful account suspensions. However, implementing these protections requires labor organizers to navigate complex workflows—manual data collection of driver earnings from in-app screenshots, error-prone translation to spreadsheets, and evidentiary disputes during arbitration. Organizers thus lack the tools to support their coordinated data work~\cite{khovanskaya2019tools} needed to contest wrongful deactivations resulting from arbitrary AI and algorithmic decisions. This creates an urgent need for designing and deploying systems guided by policy requirements~\cite{jackson2014policy, yang2024future} to support organizers.

We designed and deployed \TOOL{}, a computational tool for labor organizers, through a 6-month collaboration with the State of Washington's largest rideshare labor union (hereafter referred to as \LU{}). \TOOL{} operationalizes the state's policy mandate that platforms compensate deactivated drivers based on their 12-week pre-deactivation earnings (hereafter referred to as lost wage). The tool integrates with \LU{}'s existing workflows, syncing historical rideshare trip and earnings data and generating arbitration-ready reports to estimate lost wages for wrongfully deactivated drivers. To inform the design of \TOOL{}, we first conducted a formative study of \LU{}'s deactivation appeals process, identifying operational challenges that shaped the tool’s features. We then carried out a 3-month field deployment, resulting in 178 account sign-ups, to evaluate its real-world use and impact. In this paper, we ask:
\begin{enumerate}
\item[RQ1:] \textit{What processes, challenges, and outcomes define the deactivation appeals workflow for \LU{} staff?}
\item[RQ2:] \textit{How does the computational tool we built, \TOOL{}, improve the deactivation appeals workflow for \LU{} staff?}
\end{enumerate}

Our analysis reveals three key findings. First, \TOOL{} could reduce lost wage calculation time by over 95\% (from 25 to 2.5 minutes for gathering drivers' historical earnings, and from 2-3 hours to 2 minutes for report generation per case) and eliminated manual data entry errors, enabling the legal team to generate lost wage reports up much faster. Second, the tool provided legal teams with data access that extended beyond lost wage calculations to include independent audits and cross-referencing of platform-provided information, strengthening their position during arbitration. Third, despite these benefits, technical challenges persisted—including data synchronization issues, platform security warnings that frightened drivers, and consent process frictions—highlighting the tensions of technical solutions in high-stakes labor contexts.

Our work has broader implications for CSCW/HCI scholarship. Our findings demonstrate how tools can transform regulatory mandates into actionable evidence, developing new practices for contesting algorithmic harms through coordinated data work. We also reveal how institutional incentives and invisible maintenance labor determine tool adoption success echoing past work~\cite{orlikowski1991studying, star1999layers}. Finally, our work illustrates the ethical and practical trade-offs that emerge when tools mediate data access in information and power-asymmetric environments. In summary, our contributions are as follows:
\begin{itemize}
\item[-] Empirical insights into the deactivation appeals workflow within a labor organization (RQ1, Section \ref{sec:formative-study})
\item[-] A policy-guided computational tool that operationalizes lost wage estimation requirements, demonstrating how such tools can support organizers in contesting AI and algorithmic platform harms (Sections \ref{sec:system}, \ref{sec:field-deployment})
\item[-] An evaluation demonstrating both the benefits and challenges of the tool's adoption within an organization (RQ2, Section \ref{sec:results})
\end{itemize}

\section{Related Work}
Our paper contributes a tool supporting labor organizers in filing lost wages claims during deactivation appeals, situated within four research areas: algorithmic control mechanisms in gig work, data-driven resistance strategies, CSCW systems for organizational coordination, and HCI-policy collaboration that translates policy guidelines into design requirements. We describe them below.

\begin{figure}[htb!]
    \centering
    \includegraphics[width=\textwidth]{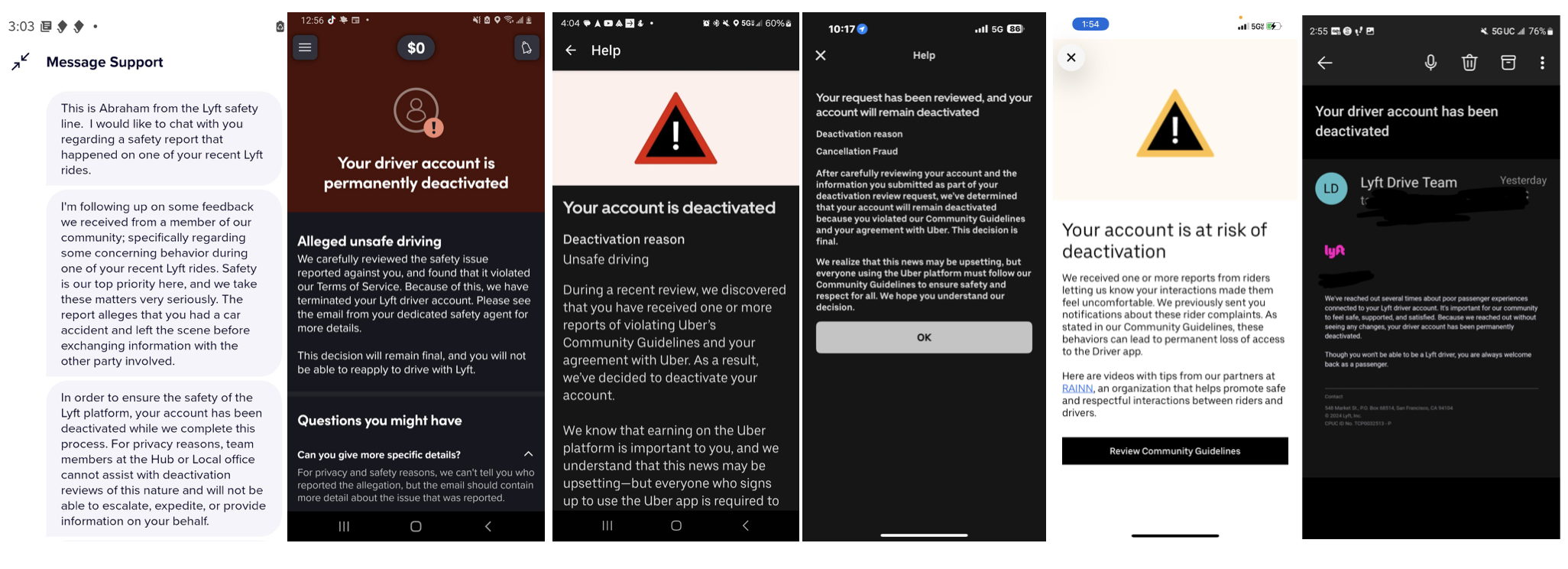}
    \caption{Screenshots of Driver Deactivation Notifications on Uber and Lyft sourced from Subreddits \texttt{r/uberdrivers} and \texttt{r/lyftdrivers}}
    \label{fig:reddit-deactivations}
\end{figure}

\subsection{Deactivation as a Mechanism for Control in the Platform Economy}

Gig economy workers are typically classified as independent contractors rather than employees under U.S labor law, which significantly impacts workplace control mechanisms~\cite{griffith2018fair}. Independent contractors should legally determine what work they do and how they perform it~\cite{IRS2025}. Platforms like Uber and DoorDash rely on ``soft control'' techniques to direct work while avoiding employer responsibilities~\cite{rosenblatAlgorithmicLaborInformation2016b}. For example, rideshare platforms ``gamify'' work with rewards designed to keep drivers online and working in specific areas—effectively controlling the work drivers do~\cite{vasudevan2022gamification,woodcockGamificationWhatIt2018}. Unlike traditional employers who can fire employees for poor performance, gig platforms must use alternative disciplinary methods because independent contractors are supposedly self-employed. Rideshare platforms employ various disciplinary strategies, including deactivation—blocking a driver's ability to work online. Deactivation is one of the most severe and destabilizing worker harms. Refer to Figure \ref{fig:reddit-deactivations} for examples of deactivation notices from rideshare platforms. 

Researchers~\cite{watkinsFaceWorkHumanCentered2023,schwartz2023deactivation,zhang2023stakeholder, calacci2022bargaining} and journalists~\cite{cbs2022deactivated,wapo2021deactivated,nyt2025deactivated} have recently studied deactivation primarily as one algorithmic management technique rather than a central issue requiring specific solutions. Early studies identified deactivation as a disciplinary threat maintaining platform control~\cite{lee2015working,rosenblatAlgorithmicLaborInformation2016b}, with drivers facing account restrictions based on acceptance rates, ratings, or alleged policy violations~\cite{cameron2022making}. These studies show how deactivation functions as algorithmic firing despite workers' supposed independence~\cite{cameron2022making,rosenblatAlgorithmicLaborInformation2016b}. However, few studies address how to remedy the financial harms of deactivation or create recourse mechanisms for drivers facing algorithmic errors or lack of due process.

The increasing platformization of work demands better understanding of how workers can contest soft control mechanisms like deactivation. Our paper examines a rare case where legal protections against unfair deactivations exist for rideshare workers in certain jurisdictions~\cite{schwartz2023deactivation}. We focus on the data work required to collect evidence for unemployment compensation, or lost wage. This legal context allows us to move beyond documenting deactivation's impacts toward creating practical tools that help drivers navigate restitution processes.

\subsection{Tools and Data Strategies for Empowering Labor Organizers}
Rideshare workers must provide documentation to prove their deactivation was unjustified, creating a significant information asymmetry between workers and platforms~\cite{ACRE2025}. Labor organizations play a crucial role in contesting deactivations through extensive data work to collect, aggregate, and utilize driver information~\cite{schwartz2023deactivation}, but platforms restrict data access by citing privacy and trade secret protections~\cite{calacci2023access}. 
To contest algorithmic management decisions like deactivations, workers need access to the very information platforms deliberately obscure.

As a result, numerous third-party tools have been developed to overcome barriers to accessing driver data. Tools like WeClock\footnote{\url{https://weclock.it/}} and ShiptCalculator\cite{calacci2022bargaining} help individual drivers collect and analyze their pay and work data to guide the development of more informed work strategies. Others like TurkOpticon~\cite{irani2013turkopticon} and Workers Info Exchange\footnote{\url{https://www.workerinfoexchange.org/}} develop tools that help individual workers aggregate their experiences to gain collective insights into working conditions. Few tools, however, center labor organizers' needs in collecting and utilizing data on behalf of the workers they represent.

The specific data needs of labor organizers include the ability to systematically collect and analyze worker payment data at scale, particularly when advocating for workers facing deactivation. This challenge is compounded by the fact that gig economy companies typically provide information in non-machine-readable formats (e.g., screenshots of receipts), making it exceptionally time and labor-intensive for organizers to process~\cite{calacci2023access}. While tools like ShiptCalculator attempt to automate data extraction using OCR, they still place the burden of collecting relevant documentation on individual workers rather than enabling authorized third parties to gather and process this information efficiently on workers' behalf~\cite{calacci2022bargaining}.

As \citet{khovanskaya2019data} notes, labor organizers have a long history of collecting and aggregating individual worker data about pay and working conditions to shape bargaining demands. However, as work has become decoupled from place~\cite{greenbaum1996back} and workers operate in increasingly geographically and informationally isolated conditions~\cite{cameron2024making}, labor organizing tactics to collect data from individual workers that historically worked well within physical factories (e.g., time and motion studies, workplace surveys) have lost some of their efficacy~\cite{khovanskaya2019tools}. Nonetheless, as \citet{woodcock2021towards} and \citet{greenbaum1996back} remind us, the information communication and cooperative work technologies that enabled the decoupling of work from place cut both ways. These same technologies can also be designed in ways that simplify sharing information about working conditions for organizing purposes.

In line with scholarship calling for the design of more worker-centered technologies that can play a \textit{``strategic role in supporting labor organizations''}~\cite{tang2023back}, we contribute a tool that supports the coordinated data work that takes place within a labor organization involved in filing a claim for financial remuneration on behalf of a deactivated rideshare worker. Our system specifically streamlines the process of collecting historical payment data, eliminating the need for labor-intensive screenshot collection and manual data processing that currently hampers organizing efforts in the gig economy. By addressing this specific gap in existing tools, our work contributes to ongoing efforts to rebalance power asymmetries in platform labor through technical interventions that support collective action.

\subsection{Understanding the Socio-Technical Challenges of Coordinated Workflows}
Despite advances in modeling workflows, interface design, and computing power, CSCW applications often fail to be adopted in practice~\cite{grudin1988cscw}. These failures can in large part be traced back to a disconnect between those who must do the work to make the system valuable and those who benefit from the system when it works well~\cite{grudin1988cscw, ackerman2000intellectual}. Human labor that sustains coordinated systems of work often goes unseen by designers and therefore, unsupported in the technical specifications~\cite{star1999layers}. Often, this can lead to what \citet{ackerman2000intellectual} calls the socio-technical gap: the disconnect between what we know we want to support socially and what ends up being supported technically. 

Overcoming the socio-technical challenges requires not just better systems design that understands \textit{all} the relevant labor that goes into animating a system, but also labor and institutional incentives to continue using the system~\cite{orlikowski1991studying}. Technology mediators who will champion novel CSCW systems and support organizational adoption are crucial in overcoming the socio-technical gap~\cite{okamuraHelpingCSCWApplications1995}. In our paper, we add to the CSCW literature on organizational adoption by presenting a field deployment of a CSCW application that aims to coordinate workflows across four different stakeholder groups within a labor union, all focused on accomplishing one collective goal: obtaining accurate lost wages calculations for deactivated workers. Our field deployment highlights the tensions that arise when creating CSCW applications for labor organizers and illuminates the workflow required to contest algorithmic harms, like deactivation.

\subsection{Integrating Policy into Tool Building}
Researchers across CSCW and HCI communities have extensively documented challenges in tech-policy collaboration, including difficulties bringing policymakers to the table \cite{spaa2019understanding} and misalignment between research methods and what policymakers need as evidence \cite{spaa2022creative}. These challenges, often framed as occurring at the boundaries of tech research and policy \cite{spaa2019understanding}, have prompted decade-long calls for increased cross-disciplinary collaboration \cite{centivany2016policy, junginger2013design, lazar2010interacting, lazar2015public, lazar2016human, urquhart2017new}. 

In response to these challenges, researchers have conceptualized the relationship between policy and technology design through various frameworks. \citet{jackson2014policy} ``Policy Knot'' conceptualizes \textit{``policy''} as a generative force that can \textit{``precede and prefigure design and practice.''} \citet{fiesler2015understanding} demonstrates how copyright law knowledge can inform online creative community design. And \citet{yang2024future} call for examining \textit{``how specific policies help shape various human-computer interactions,''} advocating for centering system-people-policy interactions in technology research.

We respond to these calls by designing a tool guided by policy implementation in the State of Washington (See Section \ref{sec:policy-context} for context on the specific policy). Our work contributes to an emerging research direction demonstrating how policy can serve as more than just tech-research’s broader impact, but as integral to its intellectual pursuits building on the rich policy-engaged research \cite{davis2012occupy, haimson2021disproportionate, rao2025rideshare, calacci2025fairfare, urquhart2022legal, disalvo2014making, whitney2021hci, blevis2007sustainable}.

\section{Background: Rideshare Regulation in the U.S.}

Policy interventions in the U.S. have begun to address harms faced by rideshare drivers, with a focus on transparency, wage protections, and due process for deactivations. In this section, we first provide background about the relevant regulatory landscape in the U.S. Then we dive deeper into the rideshare deactivation laws in the State of Washington, highlighting the key provisions which guided the development of \TOOL{}. 

\subsection{Key Rideshare Laws and Provisions in the U.S.}

While there are currently no federal laws, several state (Colorado, Minnesota, Washington) and city-level (New York City, Chicago) regulations have emerged. We summarize the laws and key provisions in Table \ref{tab:gigwork-laws}.
\setlist[itemize]{leftmargin=0.75em}
\begin{table}[h!]
\resizebox{\textwidth}{!}{
\centering
\small
\begin{tabular}{m{4cm}|m{10cm}}
\hline
\multicolumn{1}{c|}{\textbf{Law(s)}} & 
        \multicolumn{1}{c}{\textbf{Summary of Key Provisions}} \\
\hline
Colorado SB 75 (2024)\tablefootnote{\url{https://leg.colorado.gov/bills/sb24-075}} & 
\begin{itemize}
    \item Disclose to drivers the payments received from consumers and amounts paid to drivers.
    \item Develop and disclose deactivation procedures and appeals.
    \item Establish a certified driver support organization for driver deactivation representation.
    \item Periodic reporting of deactivations.
    \item Fines and civil suits for violations.
\end{itemize} \\
\hline
Minnesota HF 5247 (2024)\tablefootnote{\url{https://www.revisor.mn.gov/bills/bill.php?b=house&f=HF5247&ssn=0&y=2024}}  & 
\begin{itemize}
    \item Establishes a minimum pay rate for drivers: \$1.28 per mile, \$0.31 per minute
\end{itemize} \\
\hline
Washington HB 2076 (2022)\tablefootnote{\url{https://app.leg.wa.gov/billsummary?BillNumber=2076&Year=2021&Initiative=false}}  & 
\begin{itemize}
    \item Guarantees due process for driver deactivations through notice and appeals.
\end{itemize} \\
\hline
New York City Section 1043(b) of the NYC Charter (2016), Local Law 149, 150 (2018); Settlement (2023)\tablefootnote{\url{https://ag.ny.gov/press-release/2023/attorney-general-james-secures-328-million-uber-and-lyft-taking-earnings-drivers}, \url{https://nyc.streetsblog.org/2017/02/02/tlc-votes-to-require-uber-and-lyft-to-disclose-trip-data}, \url{https://www.nyc.gov/assets/tlc/downloads/pdf/driver_income_rules_12_04_2018.pdf}}  & 
\begin{itemize}
    \item Sets minimum wage for app-based drivers (\$17.22/hour after expenses).
    \item Wage formula based on per-minute, per-mile, and utilization rate.
    \item Requires rideshare platforms to provide detailed trip data to regulators.
\end{itemize} \\
\hline
Seattle Transportation Network Company Driver Deactivation Rights Ordinance (2020)\tablefootnote{\url{https://library.municode.com/wa/seattle/codes/municipal_code?nodeId=TIT14HURI_CH14.32TRNECODRDERI}}  & 
\begin{itemize}
    \item Guarantees protection from unwarranted deactivation
    \item Establishes a Driver Resolution Center to provide driver resolution services
    \item Enables compensation (with interest) during deactivation period
\end{itemize} \\
\hline

Chicago Transportation Network Providers Ordinance (2014)\tablefootnote{\url{https://www.chicago.gov/city/en/depts/bacp/provdrs/vehic/svcs/tnp.html}}  & 
\begin{itemize}
    \item Mandates public release of anonymized trip data for rideshare services.
\end{itemize} \\
\hline
\end{tabular}
}
\caption{Gig Worker Regulation in the United States}
\label{tab:gigwork-laws}
\end{table}

\subsection{Deactivation Policy Context in State of Washington, U.S.}
The research setting for this paper is the State of Washington in the United States. We provide more context of the policies in this state regarding the deactivation of rideshare drivers in discussion with \LU{} and drawing on the only other empirical work in partnership with \LU{} in Seattle (the largest city in the State of Washington)\cite{schwartz2023deactivation}. We also highlight specific bill language that helped guide and motivate the development of the \TOOL{}.

\label{sec:policy-context}

Before legislative reforms, rideshare platforms could deactivate drivers’ accounts algorithmically and without explanation, offering no clear appeals process or institutional support, and forcing drivers to prove their innocence while companies faced no requirement for fair investigations or consistent enforcement \cite{nyt2025deactivated, schwartz2023deactivation}. This lack of transparency left drivers vulnerable to sudden, often arbitrary deactivation based on unverified complaints or opaque AI and algorithmic decisions.

\begin{figure}[h!]
    \centering    \includegraphics[width=0.85\linewidth]{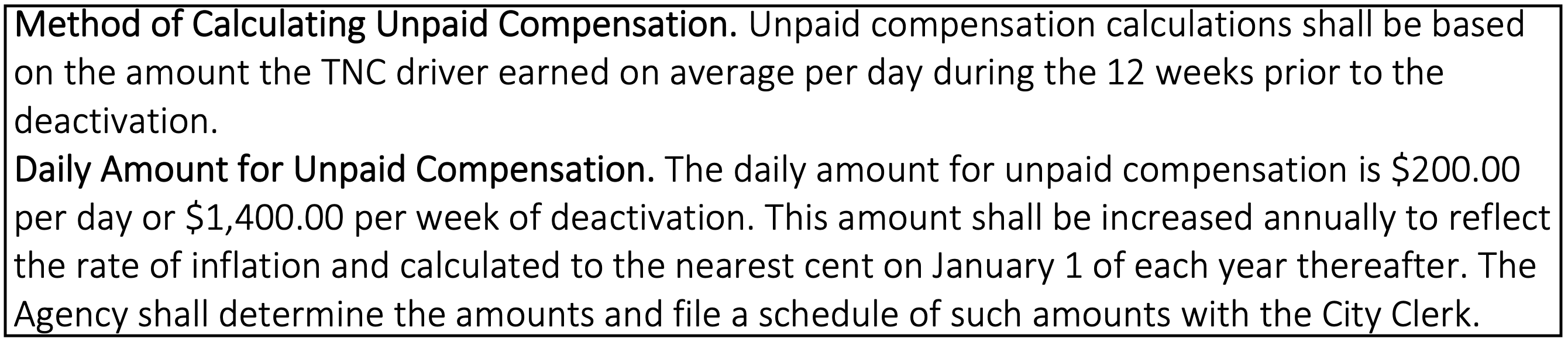}
    \caption{Lost Wage Provision From the Seattle Human Rights Rules (SHRR) Chapter 200, Rules for administering the Transportation Network Company Driver Deactivation Rights Ordinance,
Seattle Municipal Code 14.32}
    \label{fig:lost-wage-provision}
\end{figure}

\begin{figure}[h!]
    \centering    \includegraphics[width=0.85\linewidth]{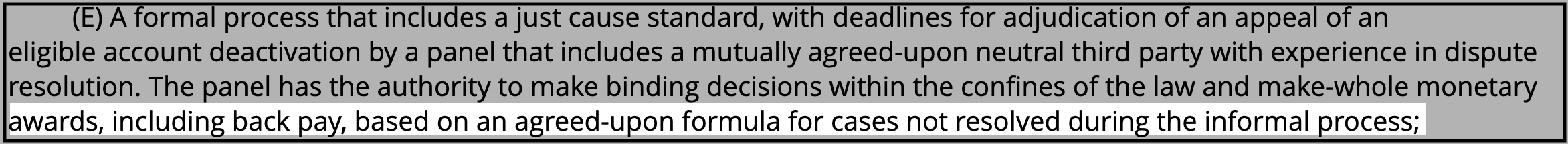}
    \caption{Lost Wage Provision From Washington House Bill 2076 codified in the Revised Code of Washington 49.46.300}
    \label{fig:lost-wage-hb2076}
\end{figure}

The regulatory landscape for rideshare drivers in State of Washington was transformed with the Seattle ordinance (the first of its kind in the U.S.) and accompanying regulations \footnote{\url{https://www.seattle.gov/documents/Departments/LaborStandards/21_0601_DRO_Chapter\%20200_Final.pdf}} passed in 2020 with an effective date of July 1, 2021. The ordinance expanded as House Bill 2076 to the whole State of Washington and was passed in June 2022; the rules\footnote{\url{https://app.leg.wa.gov/rcw/default.aspx?cite=49.46.300}}  were also subsequently passed. The ordinance ended on December 31, 2022 and any deactivations as of January 1, 2023 is processed pursuant to agreements between the rideshare platform and a driver resolution center (currently \LU{}) as outlined in HB 2076. Among many other provisions, both laws also include specific mechanisms to ensure drivers can recover lost wages due to unwarranted deactivations (See Figures \ref{fig:lost-wage-provision}, \ref{fig:lost-wage-hb2076}). Specifically, \textbf{compensation (with 12\% interest) is calculated based on the deactivated driver’s average daily earnings over the preceding 12 weeks} or a fixed amount of \$200 per day (\$1,400 per week) if the platform fails to provide drivers earnings or accept that provided as evidence by the driver resolution center during arbitration.

The introduction of these laws has significantly improved conditions for rideshare drivers. From July 2021-Dec 2023, 80\% of drivers had their deactivations overturned when they qualified for representation \cite{schwartz2023deactivation}. Since these laws were enacted, over 1,400 drivers have been reactivated through formal appeals processes facilitated by \LU{}. Drivers have recovered approximately \$3.5 million in lost earnings under these protections.

\section{Formative Study: Deactivation Workflow Appeals Process} 
\label{sec:formative-study}
We conducted a pre-deployment study with \LU{} staff combining informal conversations and a formal semi-structured interview to understand the \textit{processes, challenges, and outcomes that define the deactivation appeals workflow for \LU{}} (the state-designated driver resolution center to help drivers submit appeals and represent them). We provide an overview of the organization, our methods and outline our findings which ultimately helped inform the design of \TOOL{}.

\subsection{An Overview of Partner Organization}

We partnered with a union, \LU{}, in the State of Washington, the state's largest organization for app-based drivers, advocates for fairness, justice, and transparency in the rideshare industry. It was established in 2020 and builds on years of driver organizing. The union operates the Driver Resolution Center, funded in part by the Seattle Office of Labor Standards and the driver resource center fund. Through its efforts, drivers have secured strong protections, including safeguards against unfair deactivation, the nation’s highest statewide pay floor, paid sick leave, unemployment benefits, and access to paid family and medical leave.

\subsubsection{Staff Roles}
The driver deactivation support process involves distinct roles, each contributing to intake, investigation, advocacy, and resolution. In Table \ref{tab:roles}, we summarize the responsibilities of Field Representatives, Organizers, Paralegals, and Attorneys.
\begin{table}[h!]
\small
\centering
\begin{tabular}{l|p{10cm}}
\hline
\textbf{Role} & \textbf{Responsibilities} \\ \hline
Field Representatives & Conduct intake interviews with deactivated drivers, investigate deactivation circumstances, gather case documentation, and enroll drivers in FairFare during interviews. \\ \hline
Organizers & Oversee Field Representatives, manage operations, and lead policy advocacy efforts. \\ \hline
Paralegals & Assist in report generation and case analysis. \\ \hline
Attorneys & Draft legal briefs, generate detailed reports, negotiate settlements, and move cases through arbitration processes. \\ \hline
\end{tabular}
\caption{Roles and Responsibilities of DU Staff in Driver Deactivation Support Process}
\label{tab:roles}
\end{table}

\subsubsection{Deactivation Workflow High-level Overview:}
\begin{figure}[htb!]
    \centering    \includegraphics[width=0.65\linewidth]{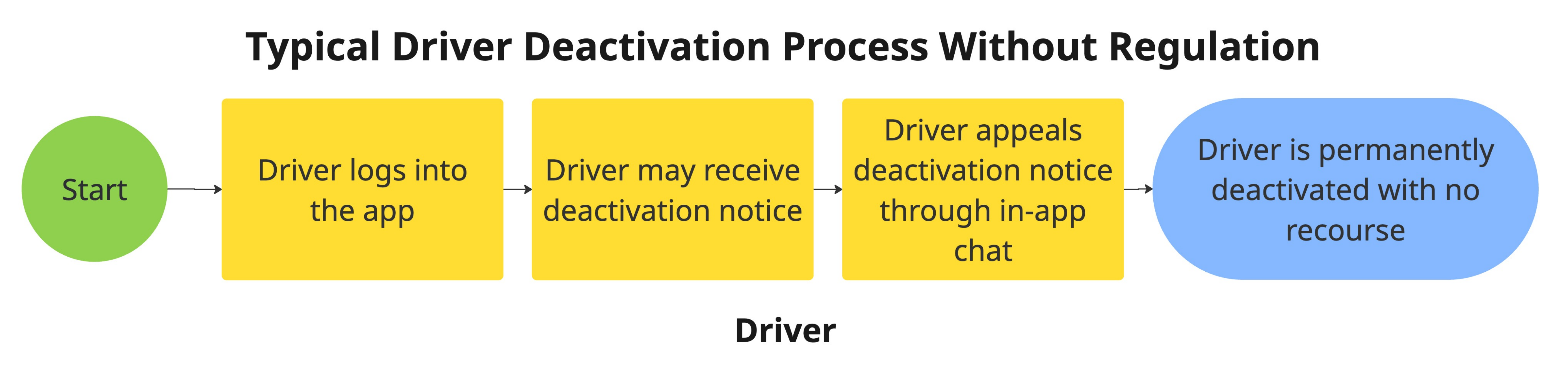}
    \caption{The \textit{typical} Driver Deactivation Process in regions without regulation establishing due process}
    \label{fig:Deactivation-process-no-regu}
\end{figure}

\begin{figure}[htb!]
    \centering
    \includegraphics[width=\textwidth]{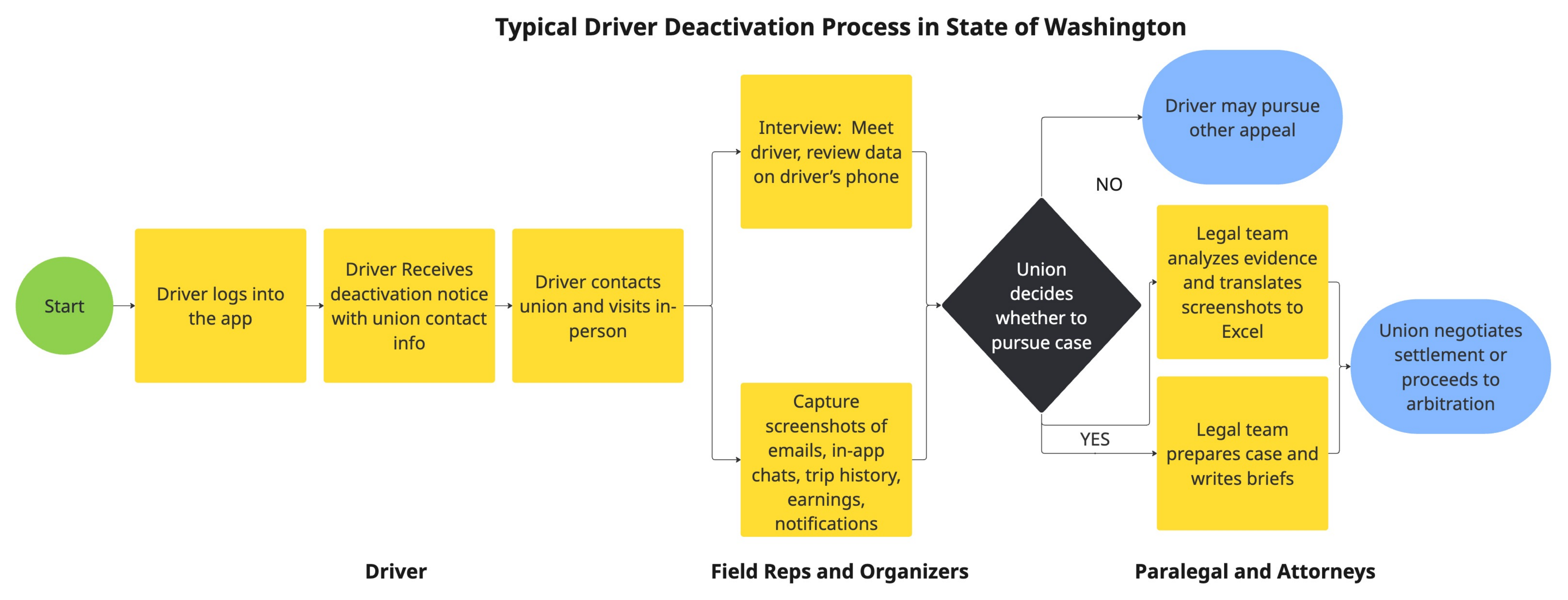}
    \caption{The Driver Deactivation Appeals Process established by \LU{} as a result of State of Washington House Bill 2076 (2022)}
    \label{fig:DU-process}
\end{figure}

In regions without any regulation, drivers often end up permanently deactivated without any ability of recourse, the typical process depicted in Figure \ref{fig:Deactivation-process-no-regu}.
As a result of regulation in the State of Washington, drivers have due process to help appeal their deactivation with representation from \LU{}. 
The high-level workflow established by \LU{} is summarized and depicted in Figure \ref{fig:DU-process}. Refer to Section \ref{sec:du-workflow} for more detailed information.

\subsection{Methods}
From June 2024 through Nov 2024, for over six months, we engaged in formative discussions via email and Zoom with \LU{} lead organizers to gain an initial understanding of their workflows and needs. Building on these insights, we conducted a virtual semi-structured interview (formative study) one month prior to tool deployment in Nov 2024 to gather more systematic data. The 75-minute interview involved two researchers as interviewers, with two professors and one engineer observing and taking notes. On \LU{}’s side, participants included two lead organizers and one lead attorney, all of whom play critical roles in managing deactivation cases.

The interview (refer Appendix \ref{app:formative-study-protocol} for the complete protocol) was structured into four stages: (1) understanding the policy context and the impact of the law on \LU{}’s work, (2) mapping the existing deactivation appeals workflow, (3) defining success metrics for appeals, and (4) identifying operational challenges. We video-recorded the session with participant consent, and transcribed the audio recording. Two authors then independently analyzed the transcripts using a top-down (deductive) thematic analysis approach, focusing on themes aligned with the interview structure. This pre-deployment study provided a detailed understanding of \LU{}’s current processes and informed subsequent tool design and deployment efforts. The study was approved by our institutions' IRB. We report our findings below focusing on the deactivation process, having already discussed the policy context in Section \ref{sec:policy-context}. 

\subsection{Findings}

\subsubsection{\textbf{The deactivation appeals process offers accessible, culturally competent support and a structured legal pathway for drivers to challenge deactivations effectively}}
\label{sec:du-workflow}
The deactivation appeals process begins when a driver is deactivated from a rideshare platform and seeks assistance. Under state law, platforms must notify drivers about their deactivation and provide contact information for \LU{}, the state-certified driver resource center. Drivers can initiate contact through multiple channels, ensuring accessibility to support services. As one participant explained: \textit{``There's a no wrong door approach for how drivers can contact the Union...folks will contact by phone call, by web form on our website, by email, and by walking into the office''}.

\begin{figure}[htb!]
    \centering
    \includegraphics[width=\linewidth]{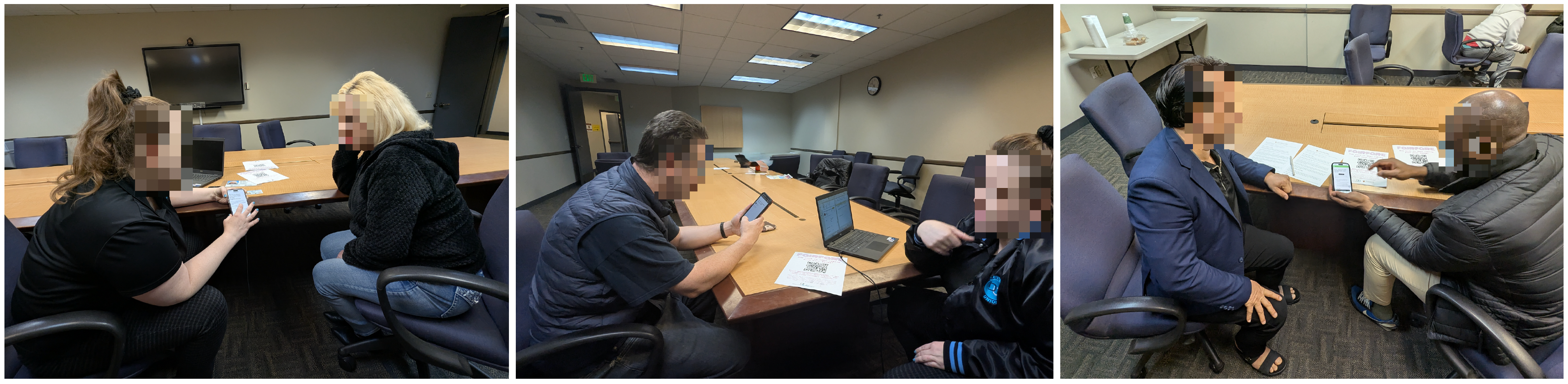}
    \caption{Field representatives schedule and conduct intake interviews up to 60 minutes with deactivated drivers to gather all evidence and then share it with the legal team for case analysis and report generation for possible settlement or arbitration with the platforms.}
    \label{fig:intake-interview}
\end{figure}

Upon initial contact, frontline staff conduct a preliminary intake to gather basic information before referring cases to specialized field representatives. This handoff is critical as it connects drivers with culturally competent support, often in their native language, given that many drivers are immigrants and non-native English speakers. The field representatives then schedule one-on-one interviews (as depicted in Figure \ref{fig:intake-interview}) to conduct thorough investigations, focusing on \textit{``the who, what, when, where, why''} of each deactivation situation as . During these interviews, representatives collect all relevant documentation from drivers as screenshots, including in-app communications, emails, earnings, and other evidence that might support their case.

The process then transitions to a more formalized legal approach once drivers sign authorization and representation forms. As one participant described: \textit{``The case is forwarded along to the legal department, which has a set of [established processes]...with an information request to the companies about why did you deactivate?''}. This creates a structured pathway where the organization can negotiate with companies on the driver's behalf, seeking either informal resolution or formal arbitration when necessary.

The entire workflow was developed through years of collaborative planning. As one participant noted: \textit{``We sat in years of meetings with drivers, with union attorneys who had experience with grievance procedures, with policy folks who had a different lens on it''}. A key design consideration was speed, with one participant characterizing their goal as establishing \textit{``a night court...a process where people can come in and out...and move people through because we knew we were going to be dealing with high [volume]''}.

\subsubsection{\textbf{Lost wage calculations rely on manual collection of drivers’ earnings data from screenshots}}
Calculating lost wages forms a critical component of the deactivation appeals process, requiring systematic data collection about driver earnings. The legal team examines the 12 weeks prior to deactivation to establish baseline earnings: \textit{``We would look at the 12 weeks prior to their deactivation and what they were making on a daily average. So added up, divided by the 84 and then use that daily average to multiply it by the amount of days they have been deactivated''}.

The primary method for gathering earnings data relies heavily on manual collection from drivers' personal records. During deactivation interviews, field representatives typically capture screenshots of earnings statements directly from drivers' phones. As one participant described: \textit{``the rep would get pictures of the drivers going to their earnings statement [on the app] and take pictures of what their weekly earnings were prior to the deactivation''}. In some cases, drivers can download PDF earnings statements if they have  access to their accounts on a desktop/laptop.

Access to earnings data becomes particularly challenging when drivers are completely locked out of their platform accounts. One representative highlighted this issue: \textit{``There's sometimes when a driver is deactivated and they're completely locked out of their app, so they're not able to go and see the 12 weeks for us to collect it...to make a demand''}. This creates significant barriers to calculating appropriate compensation claims.

In later stages of the arbitration process, especially when disputes arise over earnings calculations, the companies may provide their own records. Representatives noted that when conflicts emerge about back wages or lost earnings, \textit{``the company will provide us their...driver statements''}. This supplementary data source helps resolve discrepancies but only becomes available after cases have progressed substantially through the arbitration process.

\subsubsection{\textbf{Manual data collection for lost wages is inefficient, error-prone, and frequently requires repeated driver appointments}}
The manual nature of historical data collection creates significant inefficiencies and potential for errors throughout the appeals process. Field representatives must engage in time-consuming documentation procedures that strain resources and potentially undermine case outcomes. As one participant candidly described: 
\begin{quote}
    \textit{``The manual process of going in and gathering screenshot screenshot screenshots, to do 12 weeks...and the process you have to do to gather that manually and then to transpose the data manually—there's a lot of room for both inefficiency and error...If it's a deactivation that happened last year, and now you're backing your way through it, am I on August of last year or am I on August of this year?''}.
\end{quote}

Data entry from screenshots into calculation spreadsheets creates a particularly vulnerable point in the workflow. The paralegals or legal team must manually transcribe earnings information, dates, and other relevant details before performing calculations. One participant explained: \textit{``We gather from screenshots and then we go to an Excel sheet and we manually enter the dates, the weeks...putting that stuff, manually enter in that information and then calculate it and making sure it's right''}. They must then use separate tools to calculate interest before integrating those figures back into their demands.

When errors are discovered in the data collection phase, field representatives face the additional burden of scheduling follow-up appointments with drivers to gather missing information. One participant explained that incomplete documentation creates \textit{``a hugely inefficient process of now having to go back to gather additional documentation that was not gathered in the first place...you're talking about the driver coming back in for another appointment''}. This extends case timelines and creates additional coordination challenges for both drivers and field representatives.

\begin{figure}[htb!]
    \centering
    \includegraphics[width=\linewidth]{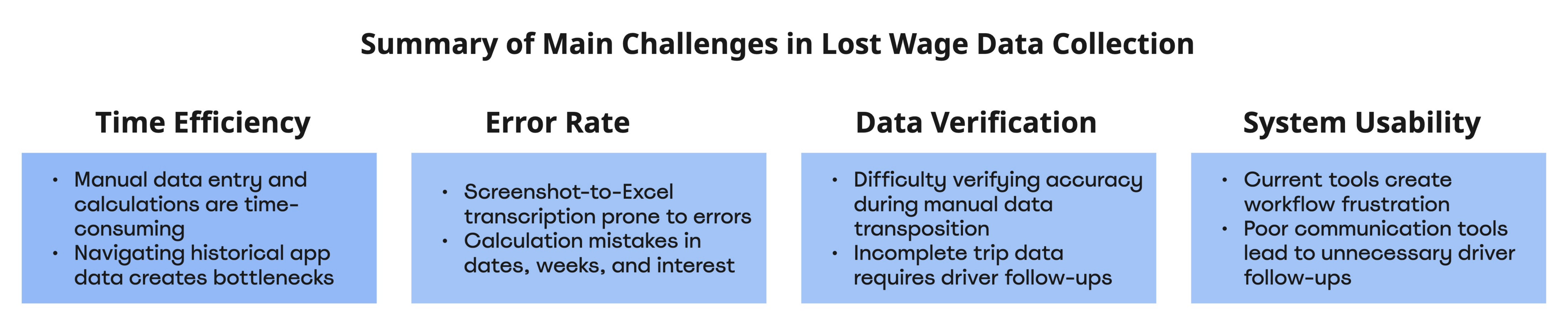}
    \caption{Summary of challenges faced by the \LU{} in lost wage data collection and analysis.}
    \label{fig:challenges}
\end{figure}

\subsubsection{\textbf{Successful resolution of a deactivation case is defined by driver satisfaction--- through reactivation, compensation or clarity}}
Successful resolution of deactivation cases takes multiple forms, with outcomes tailored to each driver's unique circumstances and preferences. Success is defined as a case being resolved to the driver's satisfaction. As one participant summarized: \textit{``A case is going to resolve with reactivation, a monetary settlement, or both''}. 

Quantitative metrics also track success and demonstrate the substantial impact of the appeals process on driver welfare. Drivers Union has achieved remarkable results, with \textit{``1,400 plus drivers reactivated''} and \textit{``more than 3.5 million in payments to drivers''} through settlement agreements.

Beyond reactivation and monetary compensation, organizers recognize that success sometimes means providing clarity and closure even when deactivation was warranted. As one participant explained: \textit{``Success can also look like for a driver who had a warranted deactivation, understanding why they were deactivated...at least understanding what it was''}.

The emotional and psychological aspects of support constitute another critical dimension of success. Organizers told us that drivers particularly value \textit{``feeling heard, understood, and respected''} through the appeals process. Organizers also noted \textit{``the stark difference in that feeling from what their interactions with support are versus what their interactions from a culturally competent in-language union representative''} who can clearly explain the situation.

Managing expectations remains essential throughout the appeals process, as outcomes are never guaranteed. Field representatives carefully communicate with drivers about realistic timelines and potential results: \textit{``We have to be very, very clear with setting expectations that there is no guarantee of the outcome''}. This expectation management continues throughout the process, particularly when large potential settlements are calculated. They also highlight that most cases will resolve before getting to the final stage through negotiated settlements that satisfy drivers' needs while avoiding the time and uncertainty of full arbitration.

\section{System: \TOOL{}---The Lost Wage Estimation Tool}
\label{sec:system}
\begin{figure}[htb!]
    \centering
    \includegraphics[width=\linewidth]{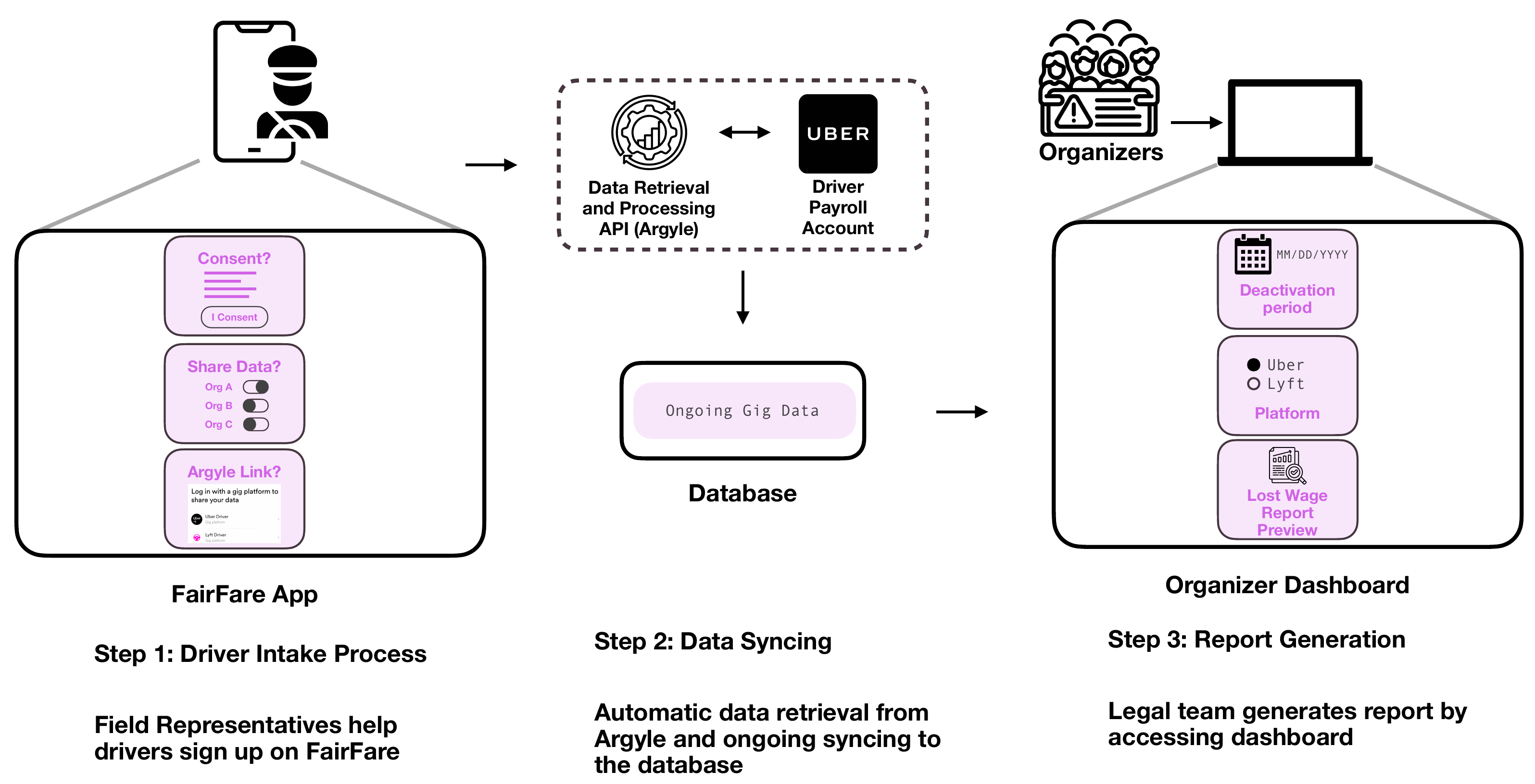}
    \caption{High Level Flow of \TOOL{}. We built \TOOL{} by extending the FairFare. Field Representatives first help drivers sign up on FairFare. Rideshare trip data then syncs from Argyle (a third party data provider) to a database. The legal team and organizers then input the drivers deactivation date, select the platform (Uber / Lyft) on a dashboard part of \TOOL{}. \TOOL{} then calculates the lost wage using the synced trip data on the database and generates a report which can be previewed or downloaded as a PDF.}
    \label{fig:tool-highlevel-flow}
\end{figure}

We developed \TOOL{} to address inefficiencies, errors, and challenges in \LU{}’s for one part of the deactivation appeals workflow --- estimation of lost wage. We aligned on this priority based on where in the deactivation appeals workflow we felt we could add the most value and \LU{} felt had the most potential for impact on their workflow. Therefore, through \TOOL{} we aimed to streamline rideshare trip data collection, improve accuracy in lost wage calculations, and reduce the manual workload for \LU{} staff by automating parts of the manual data gathering and integrating with existing workflows (See Figure \ref{fig:tool-highlevel-flow} for a high level overview). This section provides an overview of the tool’s functionality and implementation within \LU{}’s workflow.

\subsection{Key Design Criteria and Needs}
We outline the principles and functional requirements that guided the tool's development, ensuring it met the needs of \LU{} staff and integrated seamlessly into their processes. 

\noindent Through six months of conversations with \LU{} staff members, we identified \textbf{\textit{four key design principles}}:

\begin{enumerate}
    \item \textbf{Workflow Compatibility:} The tool should easily fit into staff members' existing workflow without adding significant time.
    \item \textbf{Ease of Use:} It should not require extensive technical expertise to operate; the interface and flow should remain simple.
    \item \textbf{Data Reliability:} The legal team should trust the data and reports, using them confidently as evidence during arbitration. The tool must minimize human error.
    \item \textbf{Driver Trust:} Field representatives should find it easy to explain the purpose of the tool to drivers, who are often wary of accessing their rideshare accounts and connecting it with third-party software due to fear of platform retaliation.
\end{enumerate}

\noindent In addition to these principles, we established \textbf{\textit{three key functional requirements}}:

\begin{enumerate}
    \item \textbf{Automated Data Collection:} The tool should automate historical data collection, eliminating the need for manual screenshots.
    \item \textbf{Wage Estimation:} It should automatically estimate earnings during deactivation periods using 12 weeks of historical data.
    \item \textbf{Report Generation:} The tool should generate PDF reports that include estimated lost wage and accrued interest.
\end{enumerate}

\subsection{Driver Intake Process}
\begin{figure}[htb!]
    \centering
    \includegraphics[width=\textwidth]{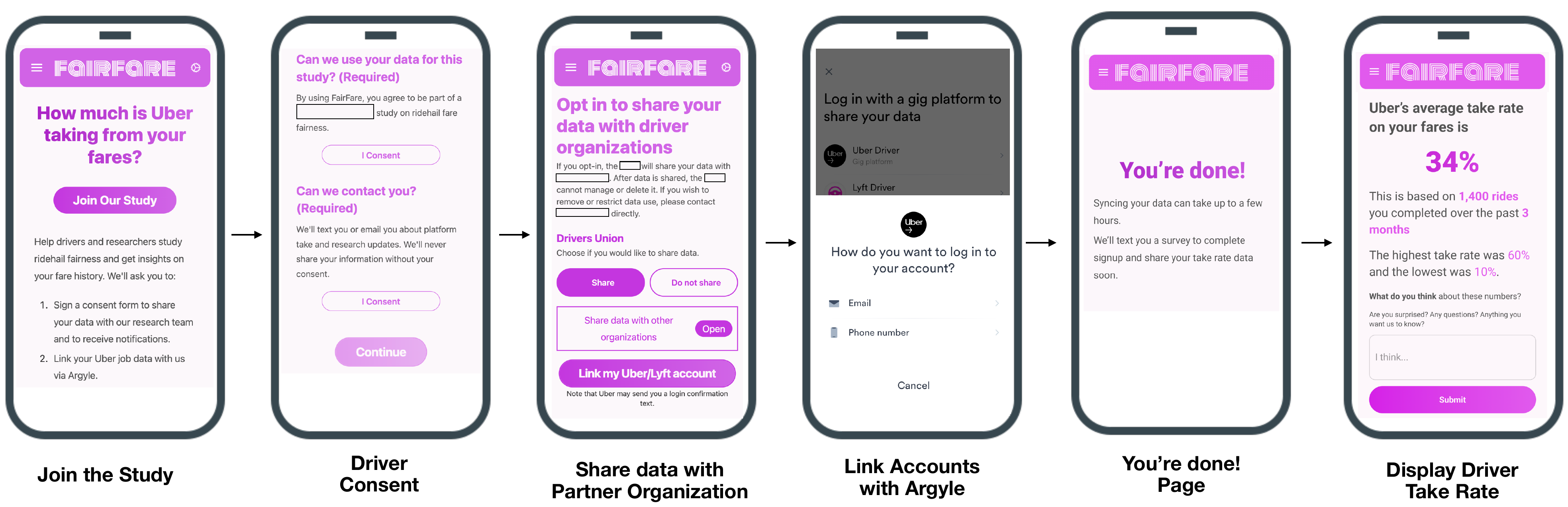}
    \caption{FairFare Driver Sign-Up Process}
    \label{fig:intake}
\end{figure}

We built \TOOL{} by extending an open-source tool called FairFare \cite{calacci2025fairfare} to sign up drivers, access their historical rideshare trip data and estimate their lost wage. FairFare already offers a straightforward driver signup process requiring minimal technical knowledge. FairFare operates through crowdsourced driver data donations of rideshare driver trip data including location, driver wages, customer charges, tips and bonuses.

FairFaire was originally built to calculate platform take rates---the percentage of rider price retained by the platform excluding tips---guided by Colorado state's SB 24-75 bill\footnote{State of Colorado Senate Bill 24-75: The bill was signed into law in June 2024, enforced since February 2025, and mandates transparency in driver wages and rider prices \url{https://leg.colorado.gov/bills/sb24-075}}. We maintained this existing flow since \LU{} appreciated the ``reward'' element that shows drivers their platform take rate after sign-up.

\noindent The driver experience consists of the following steps, as depicted in Figure \ref{fig:intake}:

\begin{enumerate}
\item Click ``Join Our Study'' on the landing page
\item Consent to participating in a university study approved by the institution's IRB
\item Opt in to share data with \LU{}
\item Link Uber or Lyft accounts (or both) by signing in with their associated phone number or email
\item View the ``You're done!'' page.
\end{enumerate}

These steps typically takes 2-3 minutes. After signup, drivers receive a text message to access their take rate once their data syncs\footnote{The tool syncs data from platforms using Argyle, a third-party service that enables access to ride data from platforms like Uber, Lyft, DoorDash and others. \url{https://argyle.com/}}. Data syncing can take 2-3 hours up to 24 hours depending on the number of driver trips. Field representatives assist drivers with FairFare signup during the intake process as one step among many others in gathering documentation.

\subsection{Lost Wage Report Generation}
\begin{figure}[htb!]
    \centering
    \includegraphics[width=\textwidth]{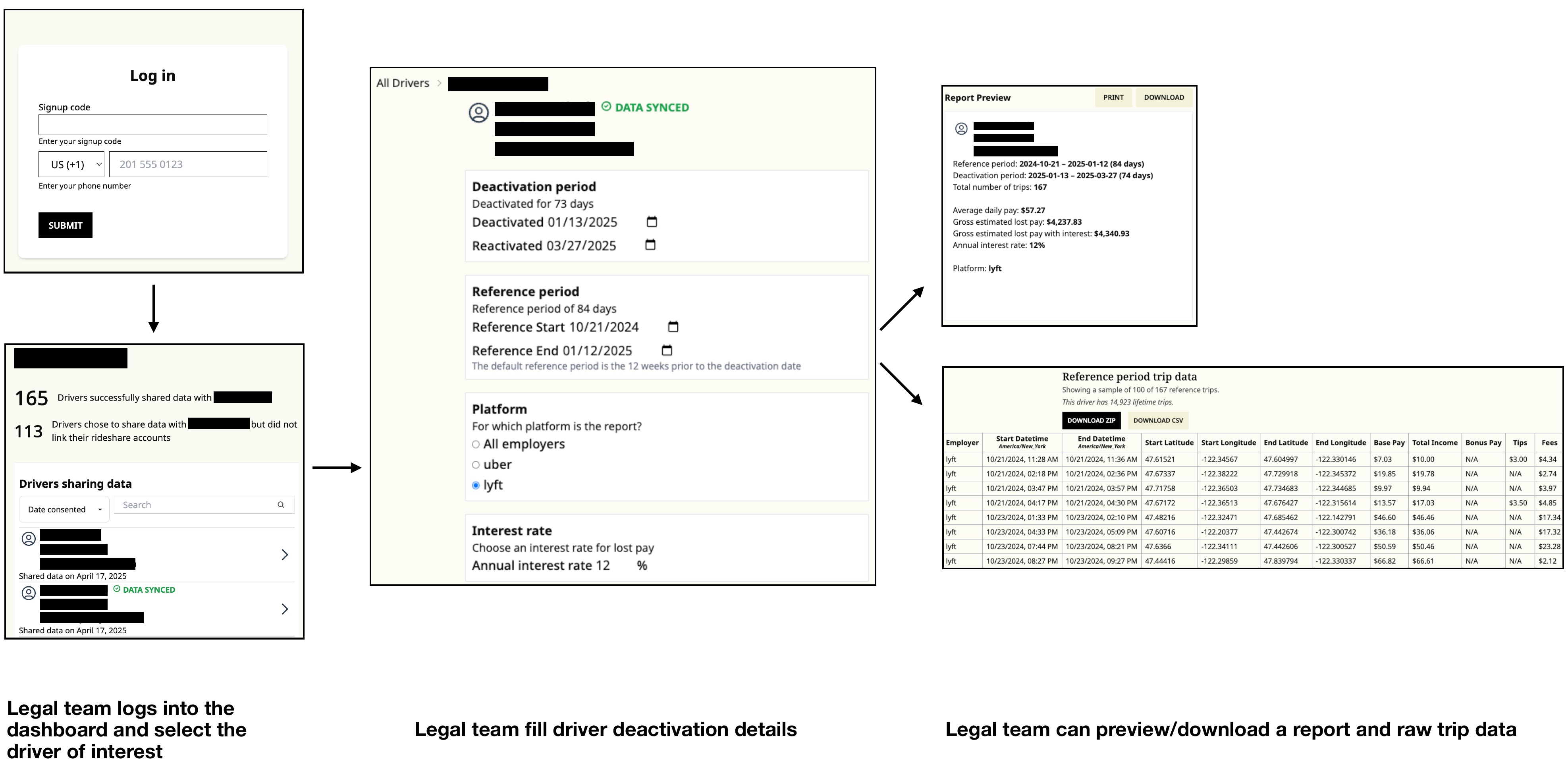}
    \caption{Lost Wage Report Generation Process. After logging into the dashboard and selecting a driver of interest, the legal team enters the date the driver has been deactivated and optionally when they are reactivated. They can then preview the lost wage report 
 and sample trip data of the selected driver. Driver PII including name, email and phone number has been redacted to preserve privacy.}
    \label{fig:report}
\end{figure}

\TOOL{} includes a backend dashboard/portal for the \LU{} legal team to access data and generate reports for deactivated drivers who have gone through \LU{}'s intake process. After data syncs, the legal team can generate lost wage reports. The legal team experience (as can be seen in Figure \ref{fig:report}) consists of the following steps:

\begin{enumerate}
\item Log into a dashboard and select the driver of interest
\item Fill driver details: enter deactivation date and optional reactivation date, choose the reference period for wage estimation (default is 12 weeks/84 days before deactivation), select the platform, and set interest rate (default is 12\%). The tool calculates average daily pay across all rides for that period and estimates lost wages during deactivation, including interest.
\item View a preview of the report and print/download a PDF
\item Preview trip data and optionally download raw data as ZIP or CSV files.
\end{enumerate}

The generated report helps provide documentation of estimated lost wage that can be submitted during arbitration proceedings.

\section{Evaluation Methods: Three Month Field Deployment and In-Person Visit to \LU{}'s Office}
\label{sec:field-deployment}
To evaluate the effectiveness and usability of the lost wage estimation tool, we conducted a three-month field deployment study, beginning in December 2024. Our main research questions for the study were:  \textit{How did \LU{} staff interact with the tool? Did it address the challenges identified during pre-deployment interviews? What is the value of the data provided by the tool, given any challenges encountered?} Specifically, we sought to determine if the tool improved \textit{efficiency} and \textit{accuracy} in the lost wage claim process while reducing errors. The tool remains in active use even as of this paper submission.

\subsection{Study Design}
\setlist[itemize]{leftmargin=0.75em}
\begin{table}[htb]
\small
\resizebox{\textwidth}{!}{%
\begin{tabular}{>{\centering\arraybackslash}m{1.5cm}|m{6cm}|m{6cm}}
\hline
\multicolumn{1}{c|}{\textbf{Metric}} & 
        \multicolumn{1}{c|}{\textbf{Aspect of challenges captured}} & 
        \multicolumn{1}{c}{\textbf{How we measured during field deployment}}  \\ \hline
\textbf{Time \newline Efficiency}
&
\begin{itemize}
\item Manual processes (e.g., entering data into Excel sheets, calculating dates/weeks/interest) were time consuming.
\item Scrolling through app history for older deactivations was inefficient.
\item High volume of cases required faster workflows.
\end{itemize}
&
\begin{itemize}
\item Time taken for intake and report generation recorded during task observations.
\item Self-reported pre-tool vs. post-tool times from the pre-study questionnaire.
\item Qualitative feedback during interviews (e.g., "How has FairFare impacted case resolution timelines?").
\end{itemize} \\ \hline

\textbf{Error Rate}
&
\begin{itemize}
\item Manual translation from screenshots to Excel sheets was error-prone. Errors occurred in calculating correct dates, weeks, and interest.
\end{itemize}
&
\begin{itemize}
\item Observations on missing fields or incorrect data entry during intake and report generation tasks.
\end{itemize} \\ \hline

\textbf{Data Verification}
&
\begin{itemize}
\item Organizers had difficulty verifying accuracy when manually transposing data.
\item Missing or incomplete trip data required follow-ups with drivers.
\end{itemize}
&
\begin{itemize}
\item Observations during report generation on whether drivers are able to verify synced trip data.
\item Follow-up interview questions about confidence in data accuracy, handling incomplete or incorrect trip data.
\end{itemize} \\ \hline

\textbf{System Usability}
&
\begin{itemize}
\item Organizers needed tools that simplified workflows and reduced frustration from manual processes.
\item Organizers wanted tools that improved communication with drivers and reduced follow-ups.
\end{itemize}
&
\begin{itemize}
\item SUS questionnaire to assess overall usability.
\item Follow-up interview questions about whether it reduced follow ups.
\end{itemize} \\

\hline
\end{tabular}
}
\caption{Evaluation Metrics and Challenges Captured}
\label{tab:eval-metrics}
\end{table}

During the deployment period, we collected data through weekly Zoom check-ins, informal conversations via email and Slack, and an in-person field visit to \LU{}’s office in March 2025. The field visit was designed to gather insights into the challenges highlighted by \LU{} organizers before deployment. Using a combination of observations, semi-structured interviews, and focus groups, we focused on four key metrics: time efficiency, error rate, data verification, and system usability as detailed in Table \ref{tab:eval-metrics}. The study was approved by our institutions' IRB.

The field visit involved detailed observations of \LU{}’s workflows and interactions with the tool. It was broadly divided into two parts based on the tool’s functionality and how \LU{} staff operate. The morning sessions focused on intake processes, including driver sign-ups and interactions with field representatives. The afternoon sessions centered on report generation workflows, involving attorneys as part of the legal team. After a driver signs up, it typically takes a few hours to up to 24 hours (if the driver has a large number of rides) for their trip data to sync to the tool. As a result, we observed report generation for drivers whose data had already synced before our visit, rather than those who signed up during the morning session.

The field visit schedule involved field observations, focus groups, and hands on technical debugging (see Appendix \ref{app:field-visit-schedule} for details). The entire protocol is present in Appendix \ref{app:field-visit-protocol}.

\subsection{Analysis}
Data collection during the field visit included detailed field notes and photographs, that were taken with participant consent and recorded in a shared document. We did not record audio or video from our field visit. Two researchers independently coded the field notes using abductive analysis thematic analysis. The researchers first sorted data from field notes into categories relating to the main challenges we identified in the pre-deployment interviews: time efficiency, error rate, data verification, and system usability. We were interested in understanding how the systems addressed each of these challenges in use. We then inductively coded the data within each category to surface how the tool addressed each challenge and identify any novel themes that arose. Subsequently, we conducted a card-sorting exercise to refine codes and identify patterns across observations, interviews, and focus groups. 

Field deployments often yield rich and diverse datasets but require significant data cleaning due to real-world complexities. In our study, we encountered similar challenges and took steps to address them. To ensure smooth operation, we organized a technical debugging session during the deployment period to resolve tool-related issues, improve communication between \LU{} staff and our research team, and establish escalation processes for ongoing support. While these activities provided valuable insights into the deployment process, they fall outside the scope of this paper. We highlight them here to remain transparent about the inevitable data cleaning and support work that accompanies field deployments.

The findings from this deployment study are based primarily on formal data collected during the field visit, supplemented by insights from ongoing informal conversations with \LU{} staff. These results provide a comprehensive evaluation of the tool’s impact on \LU{}’s workflows over three months of active use.

\section{Results}
\label{sec:results}

\begin{figure}[htb]
    \centering
    \begin{minipage}{0.76\linewidth}
        \centering
        \includegraphics[width=\linewidth]{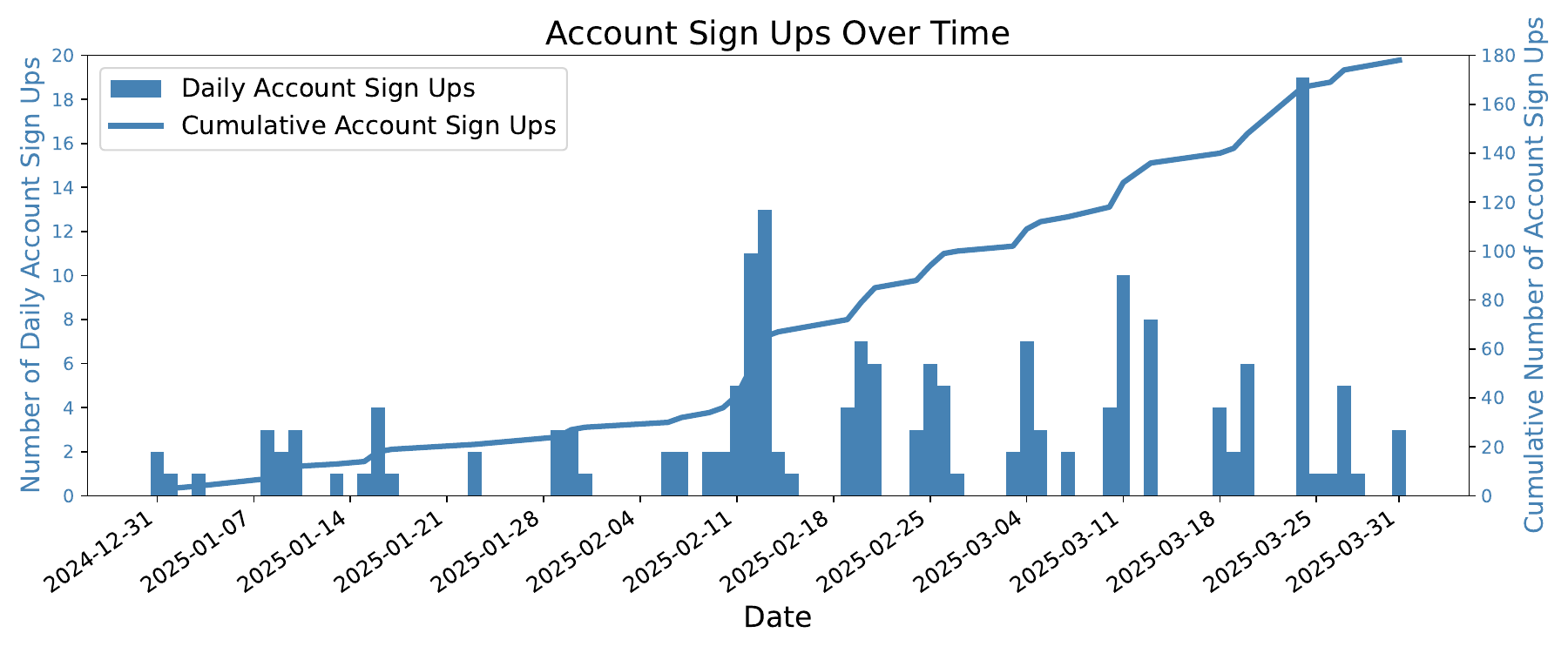}
        \caption{Account sign ups over the 3 month field deployment period from December 31st, 2024 through March 31st, 2025. We had a total of 178 unique accounts created on \TOOL{}. 143/178 accounts successfully synced their Uber or Lyft rideshare trip data.}
        \label{fig:driver-sign-ups}
    \end{minipage}%
    \hfill
    \begin{minipage}{0.18\linewidth}
        \centering
        \includegraphics[width=\linewidth]{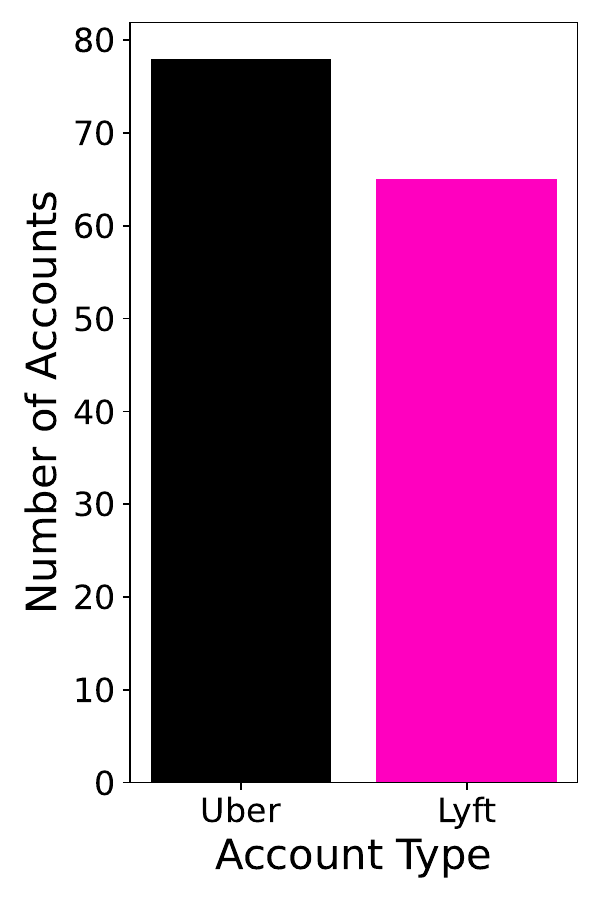}
        \caption{Distribution of the 143 accounts with successfully synced data: Uber (78) and Lyft (65)}
        \label{fig:dist-uber-lyft}
    \end{minipage}
\end{figure}

\label{sec:tool-usage}

We observed high uptake of \TOOL{} during the 3-month field deployment from December 31st, 2024 to March 31st, 2025. We recorded 122 user sign ups. And these users connected a total of 178 unique accounts. The account sign ups over time is depicted in Figure~\ref{fig:driver-sign-ups}\footnote{Our system defines a user as someone who connects their Uber or Lyft account through FairFare, and an account as either an Uber or Lyft account. Since users can be associated with multiple accounts and drivers can create multiple users, we report statistics in terms of users and accounts only.}. Of the 178 accounts, 143 successfully synced their Uber (78 accounts) or Lyft (65 accounts) rideshare trip data, while 35 accounts did not, resulting in a data sync error rate of 20\%. 

The 3-month field deployment of \TOOL{} provided valuable insights into its effectiveness and limitations when integrated into the deactivation appeals workflow. The tool streamlined key tasks, reduced data collection errors, and enhanced data access, but also introduced new socio-technical challenges. We organize our results into three themes: how the tool fit into existing workflows while not fully replacing existing ones, how it reduced errors while creating new challenges, and how its value extended beyond lost wage claims.

\subsection{\TOOL{} Made Intake and Report Generation Workflows Faster While Augmenting Existing Practices}

\textbf{\TOOL{} fit easily into field representatives' intake process, automatic specific steps, while complementing broader workflows which continued to rely on screenshots.} Field representatives integrated \TOOL{} into existing in-person interviews, typically lasting 30--60 minutes. The tool's sign-up step took only 2--3 minutes when functioning smoothly, a drastic reduction from the 20--25 minutes previously required for manual data collection with screenshots. This efficiency particularly benefited cases with extensive ride histories, where the burden of screenshotting many years of trips \footnote{Although for estimating lost wage, only 12-week pre-deactivation period is required, attorneys prefer to have access to the drivers entire ride history to make compelling legal arguments and prevent repeated driver visits.} was previously overwhelming. As one field representative noted, \textit{``This is really good work, and saves a lot of time when it works.''} Field representatives often completed the sign-up themselves on drivers' phones further streamlining the process, positioning the tool as part of a broader evidence collection effort (since they continued to collect screenshots) alongside traditional methods, rather than a replacement. The legal team encouraged collecting multiple evidence sources during in-person intake meetings to strengthen arbitration arguments; as one field representative told us: \textit{``We still collect data the same way. But it helps the legal team a lot more.''}

\textbf{For the legal team, the tool expedited report generation, cutting administrative time from hours to minutes.} After intake, \TOOL{} synchronized data within 2--24 hours, enabling attorneys to login to the dashboard and generate lost wage reports in about 2 minutes, compared to the 2--3 hours previously spent transcribing screenshots into spreadsheets and calculating interest. Attorneys then uploaded the generated lost wage reports and all supporting case documentation into their existing case management system, FileVine\footnote{\url{https://www.filevine.com/}}, where they consolidated evidence for arbitration proceedings. The automation through \TOOL{} removed a major bottleneck, especially for complex cases involving large datasets. However, attorneys wanted to trust the tool more and better understand long term patterns before fully replacing their existing screenshot-based flows: \textit{``We want to have confidence that the accounts are syncing correctly. We don’t know yet if there is a long term issue, we want to drill down a little more.''} The tool also helped the legal team minimize requests for additional data from repeated driver visits. As one attorney told us:
\begin{quote}
    \textit{``Instead of the legal team returning to the field reps for a lot of additional information and going back and forth, the data collected by the tool makes it a lot easier now to get that data without much effort.''}
\end{quote}

\textbf{Attorneys reported high usability of the tool, although findings are based on limited participants.} Using the System Usability Scale (SUS)\cite{brooke1996sus}, the two attorneys scored the tool 75 and 82.5 (See Figure \ref{fig:sus} in the Appendix), corresponding to Sauro--Lewis curved grades of B and A, indicating above-average usability. While the small sample size limits generalization, these results affirm the tool's fit for legal workflows and provide a baseline for future improvements.

\subsection{\TOOL{} Reduced Manual Errors but Surfaced New Technical and Organizational Challenges}

\begin{figure}
    \centering
    \includegraphics[width=0.4\linewidth]{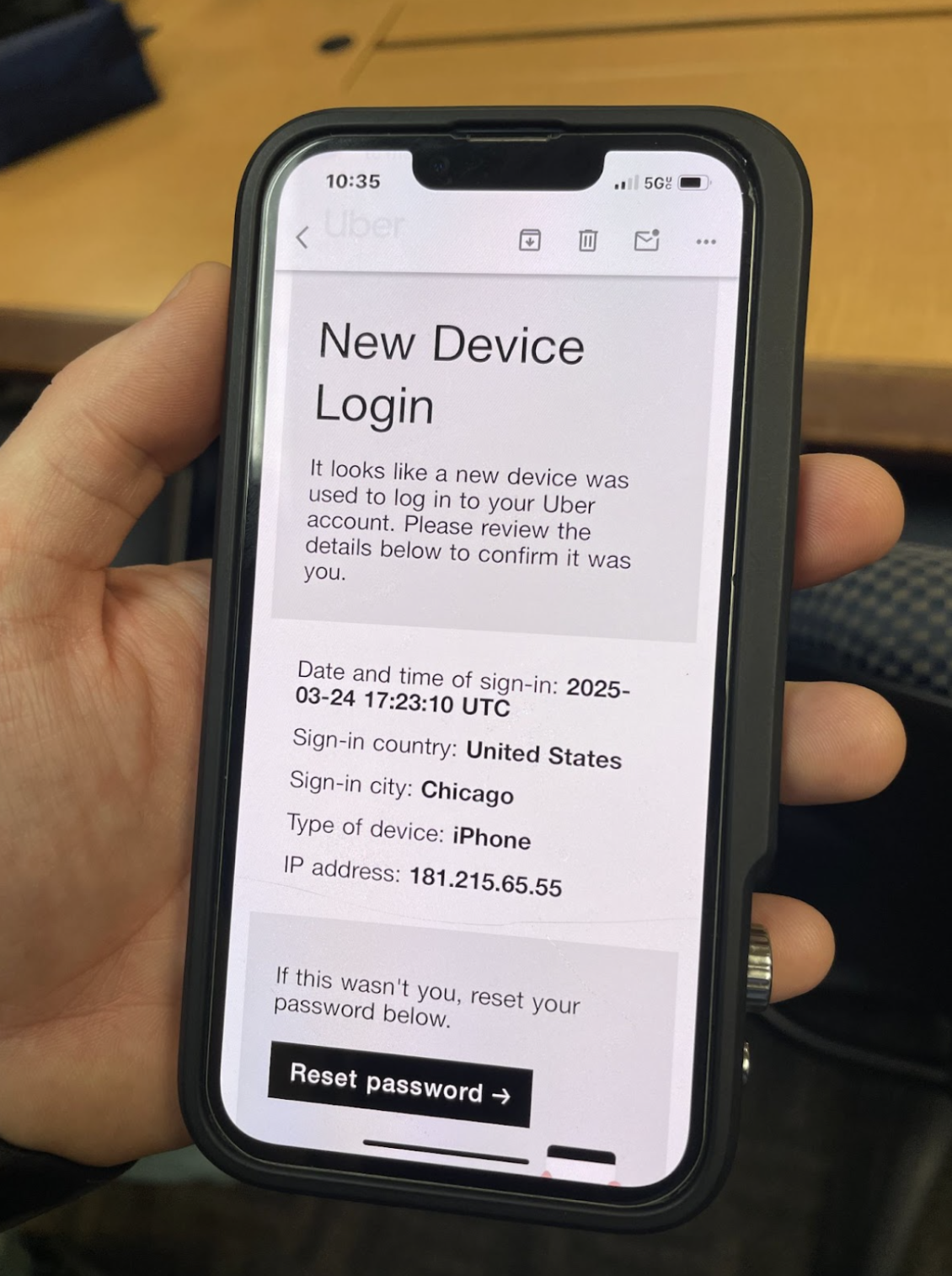}
    \caption{Occasionally, Uber's security features generate location-mismatch warnings over email during FairFare sign-up. This image of a driver's phone depicts a warning email from Uber showing Chicago instead of Seattle as the sign-up location, leading to the driver feeling confused and afraid of getting locked out of the app.}
    \label{fig:uber-location-mismatch}
\end{figure}

\textbf{\TOOL{} eliminated key data collection errors during intake and report generation processes.} Field representatives encountered fewer missing data errors since the tool automated extraction from driver accounts, addressing pre-deployment challenges: navigating multiple earnings statements, sourcing data for specific dates, and calculating interest. Similarly, attorneys no longer faced the risk of transcribing errors from screenshots into spreadsheets, streamlining case preparation stating that \textit{``this is perfect for us. When it works, it work's great!''}

\textbf{However, the tool introduced new technical obstacles that complicated intake and report generation.} During intake, challenges included weak internet connections requiring restarts, session issues necessitating browser changes, and OTP verification complications for drivers using multiple phones. Consent processes also created friction; IRB-approved language often appeared intimidating or unclear, particularly for drivers facing language barriers or where field representatives completed the process on their behalf. These tensions reflected a broader misalignment between academic consent standards and organizing realities, where field representatives carried the reputational burden of the tool. Additionally, Uber's location-mismatch warnings (Figure \ref{fig:uber-location-mismatch}) which occur during account sign up and linking on FairFare, frightened drivers, requiring reassurance from staff.

\textbf{During report generation, errors with account linking and large data syncs created frustrations that the legal team lacked capacity to resolve internally.} Approximately 20\% of accounts failed to sync data (typically Uber and not Lyft). Most sync failures occurred with Uber accounts and were typically due to errors on Argyle, which were beyond our control. Another error type arose when drivers missed consenting to data sharing with \LU{}; however, we did not systematically track statistics for this issue, as it was usually resolved after prompting the driver to re-login and provide consent. This disappointed the attorneys who told us that: \textit{``when Uber data doesn’t sync, it puts a damper on the whole case''}. These issues, while sometimes resolving over time, were confusing for the legal team, who lacked the tools to diagnose errors independently. Attorneys described situations where, after repeated failures, they abandoned the tool for certain cases. As one attorney noted, successful long-term adoption would require internal capacity akin to \textit{``an IT guy who does this''} for some of their other technical tools.

\subsection{Legal Teams Leveraged \TOOL{}'s Data Beyond Lost Wage Claims to Strengthen Arbitration and Conduct Independent Audits}

\textbf{While designed for lost wage estimation, \TOOL{} provided the legal team with strategic data advantages across other arbitration workflows too.} Attorneys told us that \TOOL{} data was \textit{``a big value add.''} Beyond calculating lost wages, attorneys used the data to initiate negotiations with platforms—such as using Uber data to open discussions with Lyft—and to cross-check company-provided information. For instance, in one case, attorneys validated a disputed Vancouver, Washington trip excluded by the company by referencing latitude and longitude data captured by the tool. For instances where data did not show up on the attorney's dashboard due to drivers forgetting to click on the consent button, they wanted us to share the data manually with them: \textit{``I was hoping you could manually pull the data and send to us when you have a chance as we have started negotiations with Lyft but they will not respond typically until we provide that information.''}

\textbf{This expanded data access helped rebalance information asymmetries and enhanced the legal team's ability to advocate effectively.} Attorneys used the comprehensive trip data to audit platform records and identify discrepancies, strengthening their position in arbitration and supporting drivers beyond the immediate lost wage claims. These practices highlight how \TOOL{} extended its value as an evidentiary resource throughout the appeals process.

\section{Discussion}

Our deployment of \TOOL{} demonstrates how policy-guided tools can empower labor organizers to contest AI and algorithmic harms, while revealing important tensions in organizational adoption and data stewardship. \TOOL{} could reduce overall documentation time by over 95\% (25 to 2.5 minutes/case for intake, and 2-3 hours to 2 minutes/case for report generation). Further, it eliminated manual data translation errors identified in pre-deployment workflows.
However, real-world implementation introduced new socio-technical challenges-from location-mismatch warnings frightening drivers (Fig.~\ref{fig:uber-location-mismatch}) to consent processes conflicting with organizers’ trust-building practices. 

These findings advance HCI and CSCW scholarship in three key areas: (1) how legislation can serve as both design constraint and opportunity for researchers and worker advocacy tool builders, (2) why institutional incentives and invisible maintenance labor determine CSCW tool adoption success, and (3) what ethical and practical trade-offs emerge when tools mediate data access in information and power-asymmetric environments. We situate these contributions within ongoing debates about rebalancing information asymmetries \cite{braverman1998labor}, bridging socio-technical gaps \cite{ackerman2000intellectual}, and preserving worker agency in data-driven organizing \cite{khovanskaya2019tools}. Finally, we reflect on limitations of our work and provide concrete recommendations for future research.

\subsection{Policy as a Design Constraint and Opportunity for Integrating Regulatory Contexts into Tool Development}
The design and deployment of \TOOL{} were uniquely shaped by explicit policy requirements in State of Washington, which provided clear technical specifications for lost wage calculation. We operationalized legislation as a technical design constraint rather than treating policy as a downstream broader impact target by directly deriving specifications from the Seattle Ordinance's \textit{``12-week average earnings''} clause. This approach shows how policy can \textit{``precede and prefigure design and practice''} \cite{jackson2014policy}, and examines \textit{``how a specific policy helps shape various human-computer interactions''} \cite{yang2024future}, offering an underexplored mechanism for HCI-policy collaboration where system logic is grounded in legislative text.

Our deployment of \TOOL{} offers a practical implementation of \citet{jackson2014policy} call to examine the co-constitution of policy, design, and practice. We demonstrate how tools can enable organizers to produce legally defensible calculations, strengthening their position in arbitration. We also show that policy’s explicit requirements—such as the 12-week earnings period—can serve as productive design constraints by removing ambiguity, risk of dispute about what evidence is acceptable and ensuring consistency in lost wage claims. Adopting this approach can enable researchers to design more tools directly guided by policy and help ensure that legal protections translate into defensible outcomes.

Finally, our deployment raises important questions about the timing and nature of HCI involvement in policy-making. Had HCI practitioners participated in the drafting or rulemaking stages, would some challenges-such as the need for technical infrastructure or the potential to require platforms to disclose earnings information directly-been anticipated or addressed differently? This suggests the need for deeper involvement of practitioners in the rulemaking process. Building on these lessons, our experience deploying \TOOL{} in the State of Washington's regulatory context also offers insights for similar efforts in other jurisdictions where rideshare regulation is emerging. As states like Colorado and Minnesota implement their own transparency and minimum pay requirements, tools that translate policy provisions into technical specifications could play a vital role in ensuring these protections are meaningfully accessible to organizers and workers.

\subsection{Aligning Incentives to Navigate Socio-Technical Gaps in Organizational Adoption of Computational Tools}
Our deployment revealed persistent socio-technical gaps when integrating \TOOL{} into \LU{}'s workflows, despite close design efforts, emphasizing that careful design alone cannot resolve all practical deployment challenges. Issues such as spotty internet, session tracking failures, location-mismatch warnings, and language barriers illustrated how well-intentioned technical systems often falter under real-world conditions. These challenges echo \citet{ackerman2000intellectual}'s insight into the divide between social requirements and technical feasibility, underscoring the need for researchers to anticipate and plan for deployment hurdles through iterative design from the outset.

Field representatives persisted through these hurdles not due to direct workflow benefits, but because the tool provided substantial value to the legal team and reduced repeat driver visits. This aligns with \citet{grudin1988cscw}'s principle that CSCW adoption requires alignment between those doing the work and those who benefit and supports \citet{orlikowski1991studying}'s emphasis on institutional incentives sustaining system use. Field representatives' willingness to troubleshoot technical issues-despite added personal effort-highlights how organizational priorities can sometimes override individual friction points during adoption. The need for technical support also became clear; as one \LU{} staff member noted, long-term adoption would require \textit{``an IT person who does this full-time,''} resonating with \citet{star1999layers}'s work on invisible maintenance labor. Our early design sessions helped surface some of these alignment opportunities, suggesting that future researchers conduct similar participatory efforts to ensure tools match institutional incentives and operational realities.

Ultimately, we felt \TOOL{} proved most effective when automating discrete tasks (data collection, wage calculations) rather than entire workflows. Although it could have reduced overall documentation time by 95\% (from 25 to 2.5 minutes per case for intake, and from 2---3 hours to 2 minutes for report generation), field representatives continued collecting screenshots to supplement arbitration claims as requested by attorneys. This augmentation approach-where technology supports rather than replaces human expertise-aligns with foundational CSCW principles for multi-stakeholder environments~\cite{suchman1987plans,horvitz1999principles,engelbart1968research, licklider1960man}, preserving organizers' critical role in interpreting evidence and building trust with drivers. Researchers can similarly use iterative participatory design sessions to identify which discrete tasks benefit from automation and which require human judgment.

\subsection{Pairing Data Access with Trust to Effectively Contest AI and Algorithmic Platform Harms}
\label{sec:consent-tension}
\TOOL{} fundamentally transformed how the legal team accessed and used driver data, enabling practices that extended beyond lost wage calculation to include independent audits, cross-referencing of platform-provided information, and more strategic negotiation. For example, attorneys used latitude/longitude records to validate disputed trips omitted from platform records, strengthening claims during arbitration. This expanded data access directly addresses longstanding information asymmetries between workers and platforms \cite{calacci2023access, rosenblatAlgorithmicLaborInformation2016b}.  providing a concrete example of \citet{braverman1998labor} argument that workers need access to the very information platforms systematically obscure. By surfacing data that platforms often conceal, \TOOL{} empowers stakeholders to negotiate more effectively and call for stronger policy protections.

However, the deployment also surfaced new tensions around data custodianship and trust. Our IRB-approved consent process, while ensuring driver agency, added friction. Drivers had to explicitly click a \textit{``Share''} button to grant data access to \LU{} (Figure \ref{fig:intake}). Several accounts part of the 20\% discussed in Section \ref{sec:tool-usage} which resulted in data sync errors were due to missed opt-in steps, leaving the legal team without usable data. Organizers thus preferred an opt-out approach where consent is provided by default, arguing that trusted field reps—not tool designers—should explain data terms to drivers, take on the reputational burden of \TOOL{} and mediate its use. This tension between academic ethics and on-the-ground organizing realities echoes \citet{khovanskaya2019tools} concern about tools unintentionally displacing worker voice. Furthermore, this shows that data access alone cannot resolve informational and power imbalances without investing in relationship building and trust. 

Looking forward, our findings suggest the need for new models of ethical infrastructure, consent and data sharing. Future iterations might treat \TOOL{} as a service allowing organizers to adopt consent models fitting their context, while enabling secondary research with academics through data-sharing agreements. This would preserve ethical rigor while reducing barriers to adoption in high-stakes organizing settings. We call for more work to better understand the trade-offs involved and design mechanisms which better balance ethics with practical challenges.

\subsection{Limitations and Future Work}
\label{sec:limitations}

Our study has several limitations that can inform opportunities for future research and tool development.

\noindent \textbf{\textit{Limited evaluation period captures only short-term adoption:}}
Our study reports on the initial three-month field deployment and does not account for longer-term patterns of adoption, tool maintenance, or shifts in organizational practices. Future work can examine how tools like \TOOL{} evolve within organizations over extended periods, including whether they become institutionalized, require sustained technical support, or face abandonment.

\noindent \textbf{\textit{Deployment lacked internal capacity for handling technical errors:}}
Despite design, \TOOL{} introduced technical failures—including sync errors and account linking problems—that staff could not resolve independently. Future work can explore how to build organizational capacity for troubleshooting and maintaining such tools, including models where unions have dedicated technical roles or infrastructure.

\noindent \textbf{\textit{HCI researchers' role in rulemaking remains undefined:}}
Our deployment surfaced gaps where earlier HCI involvement in rulemaking might have anticipated technical infrastructure needs or consent challenges. Future work should investigate at what phase HCI researchers can engage in the policy process, how such collaborations might be structured, and what roles researchers should play—whether contributing to bill language drafting, offering technical guidance during rulemaking, or only providing post-enactment support to ensure laws translate into usable systems.

\noindent \textbf{\textit{Driver perspectives are outside the scope of this study:}}
This paper centers on labor organizers' workflows and legal teams—the primary users of \TOOL{}. While driver perspectives on data access, trust, and appeals are critical, capturing their voices deserves a dedicated longitudinal inquiry. We see this as rich ground for future stand-alone work. Our focus here, on organizers, enables us to analyze how institutional actors operationalize data-driven tools within the constraints of policy and organizational workflows—an essential but separate layer from drivers’ individual experiences, which many past works have already begun to capture \cite{rao2025rideshare, zhang2023stakeholder, rosenblatAlgorithmicLaborInformation2016b}.

\section{Conclusion}
This paper presents \TOOL{}, a tool designed with the labor union \LU{} to support lost wage claims in rideshare driver deactivation cases. During a 3-month field deployment with 178 account signups, \TOOL{} reduced time spent on specific tasks—screenshot gathering during driver intake (from 25 to 2–3 minutes) and report generation (from 2–3 hours to 2 minutes per case)—while continuing to complement broader documentation workflows that organizers maintained alongside the tool. It eliminated manual data entry errors and gave legal teams access to structured trip data they previously lacked. Organizers used this data not only to calculate lost wages but also to audit platform-provided records and strengthen arbitration cases.

Despite these benefits and close collaboration with \LU{}, the tool introduced new challenges, including data sync errors, consent frictions, and a need for ongoing technical support. While \LU{} staff continued using the tool due to its legal utility, they did not fully replace manual methods.

Our deployment shows that policy-driven tools can improve legal workflows, but their success depends on integration into existing practices, institutional incentives for sustained use, and in-house technical capacity. Our findings also raise questions for future work: how might HCI researchers engage earlier in the rule writing process, and what new consent models can balance ethical rigor with organizing realities?

\section*{Acknowledgements}
We thank our partners at \LU{} for all their support during design and deployment of \TOOL{}. The project is partially funded through the Mozilla Technology Fund (2023). Varun was partly supported by a Google Policy Fellowship and a Data Driven Social Science Fellowship at Princeton University.

\bibliographystyle{ACM-Reference-Format}
\bibliography{references}

%%% -*-BibTeX-*-
%%% Do NOT edit. File created by BibTeX with style
%%% ACM-Reference-Format-Journals [18-Jan-2012].

\begin{thebibliography}{54}

%%% ====================================================================
%%% NOTE TO THE USER: you can override these defaults by providing
%%% customized versions of any of these macros before the \bibliography
%%% command.  Each of them MUST provide its own final punctuation,
%%% except for \shownote{} and \showURL{}.  The latter two
%%% do not use final punctuation, in order to avoid confusing it with
%%% the Web address.
%%%
%%% To suppress output of a particular field, define its macro to expand
%%% to an empty string, or better, \unskip, like this:
%%%
%%% \newcommand{\showURL}[1]{\unskip}   % LaTeX syntax
%%%
%%% \def \showURL #1{\unskip}           % plain TeX syntax
%%%
%%% ====================================================================

\ifx \showCODEN    \undefined \def \showCODEN     #1{\unskip}     \fi
\ifx \showISBNx    \undefined \def \showISBNx     #1{\unskip}     \fi
\ifx \showISBNxiii \undefined \def \showISBNxiii  #1{\unskip}     \fi
\ifx \showISSN     \undefined \def \showISSN      #1{\unskip}     \fi
\ifx \showLCCN     \undefined \def \showLCCN      #1{\unskip}     \fi
\ifx \shownote     \undefined \def \shownote      #1{#1}          \fi
\ifx \showarticletitle \undefined \def \showarticletitle #1{#1}   \fi
\ifx \showURL      \undefined \def \showURL       {\relax}        \fi
% The following commands are used for tagged output and should be
% invisible to TeX
\providecommand\bibfield[2]{#2}
\providecommand\bibinfo[2]{#2}
\providecommand\natexlab[1]{#1}
\providecommand\showeprint[2][]{arXiv:#2}

\bibitem[Ackerman(2000)]%
        {ackerman2000intellectual}
\bibfield{author}{\bibinfo{person}{Mark~S Ackerman}.} \bibinfo{year}{2000}\natexlab{}.
\newblock \showarticletitle{The intellectual challenge of CSCW: the gap between social requirements and technical feasibility}.
\newblock \bibinfo{journal}{\emph{Human--Computer Interaction}} \bibinfo{volume}{15}, \bibinfo{number}{2-3} (\bibinfo{year}{2000}), \bibinfo{pages}{179--203}.
\newblock


\bibitem[{Action Center on Race and the Economy}(2025)]%
        {ACRE2025}
\bibfield{author}{\bibinfo{person}{{Action Center on Race and the Economy}}.} \bibinfo{year}{2025}\natexlab{}.
\newblock \bibinfo{title}{Driven Out By {AI}: How {Uber} Deactivations Force Drivers into Chatbot Hell and Financial Crisis}.
\newblock
\urldef\tempurl%
\url{https://acrecampaigns.org/research_post/driven-out-by-ai/}
\showURL{%
\tempurl}
\newblock
\shownote{Accessed: May 3, 2025}.


\bibitem[Blevis(2007)]%
        {blevis2007sustainable}
\bibfield{author}{\bibinfo{person}{Eli Blevis}.} \bibinfo{year}{2007}\natexlab{}.
\newblock \showarticletitle{Sustainable interaction design: invention \& disposal, renewal \& reuse}. In \bibinfo{booktitle}{\emph{Proceedings of the SIGCHI conference on Human factors in computing systems}}. \bibinfo{pages}{503--512}.
\newblock


\bibitem[Braverman(1998)]%
        {braverman1998labor}
\bibfield{author}{\bibinfo{person}{Harry Braverman}.} \bibinfo{year}{1998}\natexlab{}.
\newblock \bibinfo{booktitle}{\emph{Labor and monopoly capital: The degradation of work in the twentieth century}}.
\newblock \bibinfo{publisher}{nyu Press}.
\newblock


\bibitem[Brooke et~al\mbox{.}(1996)]%
        {brooke1996sus}
\bibfield{author}{\bibinfo{person}{John Brooke} {et~al\mbox{.}}} \bibinfo{year}{1996}\natexlab{}.
\newblock \showarticletitle{SUS-A quick and dirty usability scale}.
\newblock \bibinfo{journal}{\emph{Usability evaluation in industry}} \bibinfo{volume}{189}, \bibinfo{number}{194} (\bibinfo{year}{1996}), \bibinfo{pages}{4--7}.
\newblock


\bibitem[Calacci et~al\mbox{.}(2025)]%
        {calacci2025fairfare}
\bibfield{author}{\bibinfo{person}{Dana Calacci}, \bibinfo{person}{Varun Nagaraj~Rao}, \bibinfo{person}{Samantha Dalal}, \bibinfo{person}{Catherine Di}, \bibinfo{person}{Kok-Wei Pua}, \bibinfo{person}{Andrew Schwartz}, \bibinfo{person}{Danny Spitzberg}, {and} \bibinfo{person}{Andr{\'e}s Monroy-Hern{\'a}ndez}.} \bibinfo{year}{2025}\natexlab{}.
\newblock \showarticletitle{FairFare: A Tool for Crowdsourcing Rideshare Data to Empower Labor Organizers}.
\newblock \bibinfo{journal}{\emph{arXiv preprint arXiv:2502.11273}} (\bibinfo{year}{2025}).
\newblock


\bibitem[Calacci and Pentland(2022)]%
        {calacci2022bargaining}
\bibfield{author}{\bibinfo{person}{Dan Calacci} {and} \bibinfo{person}{Alex Pentland}.} \bibinfo{year}{2022}\natexlab{}.
\newblock \showarticletitle{Bargaining with the black-box: Designing and deploying worker-centric tools to audit algorithmic management}.
\newblock \bibinfo{journal}{\emph{Proceedings of the ACM on Human-Computer Interaction}} \bibinfo{volume}{6}, \bibinfo{number}{CSCW2} (\bibinfo{year}{2022}), \bibinfo{pages}{1--24}.
\newblock


\bibitem[Calacci and Stein(2023)]%
        {calacci2023access}
\bibfield{author}{\bibinfo{person}{Dan Calacci} {and} \bibinfo{person}{Jake Stein}.} \bibinfo{year}{2023}\natexlab{}.
\newblock \showarticletitle{From access to understanding: Collective data governance for workers}.
\newblock \bibinfo{journal}{\emph{European Labour Law Journal}} \bibinfo{volume}{14}, \bibinfo{number}{2} (\bibinfo{year}{2023}), \bibinfo{pages}{253--282}.
\newblock


\bibitem[Cameron(2022)]%
        {cameron2022making}
\bibfield{author}{\bibinfo{person}{Lindsey~D Cameron}.} \bibinfo{year}{2022}\natexlab{}.
\newblock \showarticletitle{“Making out” while driving: Relational and efficiency games in the gig economy}.
\newblock \bibinfo{journal}{\emph{Organization Science}} \bibinfo{volume}{33}, \bibinfo{number}{1} (\bibinfo{year}{2022}), \bibinfo{pages}{231--252}.
\newblock


\bibitem[Cameron(2024)]%
        {cameron2024making}
\bibfield{author}{\bibinfo{person}{Lindsey~D Cameron}.} \bibinfo{year}{2024}\natexlab{}.
\newblock \showarticletitle{The making of the “good bad” job: How algorithmic management manufactures consent through constant and confined choices}.
\newblock \bibinfo{journal}{\emph{Administrative Science Quarterly}} \bibinfo{volume}{69}, \bibinfo{number}{2} (\bibinfo{year}{2024}), \bibinfo{pages}{458--514}.
\newblock


\bibitem[Centivany(2016)]%
        {centivany2016policy}
\bibfield{author}{\bibinfo{person}{Alissa Centivany}.} \bibinfo{year}{2016}\natexlab{}.
\newblock \showarticletitle{Policy as embedded generativity: A case study of the emergence and evolution of hathiTrust}. In \bibinfo{booktitle}{\emph{Proceedings of the 19th ACM conference on computer-supported cooperative work \& social computing}}. \bibinfo{pages}{926--940}.
\newblock


\bibitem[Davis et~al\mbox{.}(2012)]%
        {davis2012occupy}
\bibfield{author}{\bibinfo{person}{Janet Davis}, \bibinfo{person}{Harry Hochheiser}, \bibinfo{person}{Juan~Pablo Hourcade}, \bibinfo{person}{Jeff Johnson}, \bibinfo{person}{Lisa Nathan}, {and} \bibinfo{person}{Janice Tsai}.} \bibinfo{year}{2012}\natexlab{}.
\newblock \showarticletitle{Occupy CHI! engaging US policymakers}.
\newblock In \bibinfo{booktitle}{\emph{CHI'12 Extended Abstracts on Human Factors in Computing Systems}}. \bibinfo{pages}{1139--1142}.
\newblock


\bibitem[DiSalvo et~al\mbox{.}(2014)]%
        {disalvo2014making}
\bibfield{author}{\bibinfo{person}{Carl DiSalvo}, \bibinfo{person}{Jonathan Lukens}, \bibinfo{person}{Thomas Lodato}, \bibinfo{person}{Tom Jenkins}, {and} \bibinfo{person}{Tanyoung Kim}.} \bibinfo{year}{2014}\natexlab{}.
\newblock \showarticletitle{Making public things: how HCI design can express matters of concern}. In \bibinfo{booktitle}{\emph{Proceedings of the SIGCHI Conference on Human factors in Computing Systems}}. \bibinfo{pages}{2397--2406}.
\newblock


\bibitem[Engelbart and English(1968)]%
        {engelbart1968research}
\bibfield{author}{\bibinfo{person}{Douglas~C Engelbart} {and} \bibinfo{person}{William~K English}.} \bibinfo{year}{1968}\natexlab{}.
\newblock \showarticletitle{A research center for augmenting human intellect}. In \bibinfo{booktitle}{\emph{Proceedings of the December 9-11, 1968, fall joint computer conference, part I}}. \bibinfo{pages}{395--410}.
\newblock


\bibitem[Fiesler et~al\mbox{.}(2015)]%
        {fiesler2015understanding}
\bibfield{author}{\bibinfo{person}{Casey Fiesler}, \bibinfo{person}{Jessica~L Feuston}, {and} \bibinfo{person}{Amy~S Bruckman}.} \bibinfo{year}{2015}\natexlab{}.
\newblock \showarticletitle{Understanding copyright law in online creative communities}. In \bibinfo{booktitle}{\emph{Proceedings of the 18th ACM conference on computer supported cooperative work \& social computing}}. \bibinfo{pages}{116--129}.
\newblock


\bibitem[Greenbaum(1996)]%
        {greenbaum1996back}
\bibfield{author}{\bibinfo{person}{Joan Greenbaum}.} \bibinfo{year}{1996}\natexlab{}.
\newblock \showarticletitle{Back to Labor: Returning to labor process discussions in the study of work}. In \bibinfo{booktitle}{\emph{Proceedings of the 1996 ACM conference on Computer supported cooperative work}}. \bibinfo{pages}{229--237}.
\newblock


\bibitem[Griffith(2018)]%
        {griffith2018fair}
\bibfield{author}{\bibinfo{person}{Kati~L Griffith}.} \bibinfo{year}{2018}\natexlab{}.
\newblock \showarticletitle{The Fair Labor Standards Act at 80: Everything Old Is New Again}.
\newblock \bibinfo{journal}{\emph{Cornell L. Rev.}}  \bibinfo{volume}{104} (\bibinfo{year}{2018}), \bibinfo{pages}{557}.
\newblock


\bibitem[Grudin(1988)]%
        {grudin1988cscw}
\bibfield{author}{\bibinfo{person}{Jonathan Grudin}.} \bibinfo{year}{1988}\natexlab{}.
\newblock \showarticletitle{Why CSCW applications fail: problems in the design and evaluationof organizational interfaces}. In \bibinfo{booktitle}{\emph{Proceedings of the 1988 ACM conference on Computer-supported cooperative work}}. \bibinfo{pages}{85--93}.
\newblock


\bibitem[Haimson et~al\mbox{.}(2021)]%
        {haimson2021disproportionate}
\bibfield{author}{\bibinfo{person}{Oliver~L Haimson}, \bibinfo{person}{Daniel Delmonaco}, \bibinfo{person}{Peipei Nie}, {and} \bibinfo{person}{Andrea Wegner}.} \bibinfo{year}{2021}\natexlab{}.
\newblock \showarticletitle{Disproportionate removals and differing content moderation experiences for conservative, transgender, and black social media users: Marginalization and moderation gray areas}.
\newblock \bibinfo{journal}{\emph{Proceedings of the ACM on Human-Computer Interaction}} \bibinfo{volume}{5}, \bibinfo{number}{CSCW2} (\bibinfo{year}{2021}), \bibinfo{pages}{1--35}.
\newblock


\bibitem[Horvitz(1999)]%
        {horvitz1999principles}
\bibfield{author}{\bibinfo{person}{Eric Horvitz}.} \bibinfo{year}{1999}\natexlab{}.
\newblock \showarticletitle{Principles of mixed-initiative user interfaces}. In \bibinfo{booktitle}{\emph{Proceedings of the SIGCHI conference on Human Factors in Computing Systems}}. \bibinfo{pages}{159--166}.
\newblock


\bibitem[Irani and Silberman(2013)]%
        {irani2013turkopticon}
\bibfield{author}{\bibinfo{person}{Lilly~C Irani} {and} \bibinfo{person}{M~Six Silberman}.} \bibinfo{year}{2013}\natexlab{}.
\newblock \showarticletitle{Turkopticon: Interrupting worker invisibility in amazon mechanical turk}. In \bibinfo{booktitle}{\emph{Proceedings of the SIGCHI conference on human factors in computing systems}}. \bibinfo{pages}{611--620}.
\newblock


\bibitem[Jackson et~al\mbox{.}(2014)]%
        {jackson2014policy}
\bibfield{author}{\bibinfo{person}{Steven~J Jackson}, \bibinfo{person}{Tarleton Gillespie}, {and} \bibinfo{person}{Sandy Payette}.} \bibinfo{year}{2014}\natexlab{}.
\newblock \showarticletitle{The policy knot: Re-integrating policy, practice and design in CSCW studies of social computing}. In \bibinfo{booktitle}{\emph{Proceedings of the 17th ACM conference on Computer supported cooperative work \& social computing}}. \bibinfo{pages}{588--602}.
\newblock


\bibitem[Junginger(2013)]%
        {junginger2013design}
\bibfield{author}{\bibinfo{person}{Sabine Junginger}.} \bibinfo{year}{2013}\natexlab{}.
\newblock \showarticletitle{Design and innovation in the public sector: Matters of design in policy-making and policy implementation}.
\newblock \bibinfo{journal}{\emph{Annual Review of Policy Design}} \bibinfo{volume}{1}, \bibinfo{number}{1} (\bibinfo{year}{2013}), \bibinfo{pages}{1--11}.
\newblock


\bibitem[Keppler(2025)]%
        {nyt2025deactivated}
\bibfield{author}{\bibinfo{person}{Nick Keppler}.} \bibinfo{year}{2025}\natexlab{}.
\newblock \bibinfo{title}{They Were Deactivated From Delivering. Their Finances Were Devastated.}
\newblock
\urldef\tempurl%
\url{https://www.nytimes.com/2025/03/29/business/uber-lyft-doordash-deactivation.html}
\showURL{%
\tempurl}


\bibitem[Khovanskaya et~al\mbox{.}(2019)]%
        {khovanskaya2019tools}
\bibfield{author}{\bibinfo{person}{Vera Khovanskaya}, \bibinfo{person}{Lynn Dombrowski}, \bibinfo{person}{Jeffrey Rzeszotarski}, {and} \bibinfo{person}{Phoebe Sengers}.} \bibinfo{year}{2019}\natexlab{}.
\newblock \showarticletitle{The tools of management: Adapting historical union tactics to platform-mediated labor}.
\newblock \bibinfo{journal}{\emph{Proceedings of the ACM on Human-Computer Interaction}} \bibinfo{volume}{3}, \bibinfo{number}{CSCW} (\bibinfo{year}{2019}), \bibinfo{pages}{1--22}.
\newblock


\bibitem[Khovanskaya and Sengers(2019)]%
        {khovanskaya2019data}
\bibfield{author}{\bibinfo{person}{Vera Khovanskaya} {and} \bibinfo{person}{Phoebe Sengers}.} \bibinfo{year}{2019}\natexlab{}.
\newblock \showarticletitle{Data rhetoric and uneasy alliances: Data advocacy in US labor history}. In \bibinfo{booktitle}{\emph{Proceedings of the 2019 on Designing Interactive Systems Conference}}. \bibinfo{pages}{1391--1403}.
\newblock


\bibitem[Lazar(2010)]%
        {lazar2010interacting}
\bibfield{author}{\bibinfo{person}{Jonathan Lazar}.} \bibinfo{year}{2010}\natexlab{}.
\newblock \showarticletitle{Interacting with public policy interacting with public policy}.
\newblock \bibinfo{journal}{\emph{Interactions}} \bibinfo{volume}{17}, \bibinfo{number}{1} (\bibinfo{year}{2010}), \bibinfo{pages}{40--43}.
\newblock


\bibitem[Lazar(2015)]%
        {lazar2015public}
\bibfield{author}{\bibinfo{person}{Jonathan Lazar}.} \bibinfo{year}{2015}\natexlab{}.
\newblock \showarticletitle{Public policy and HCI: making an impact in the future}.
\newblock \bibinfo{journal}{\emph{Interactions}} \bibinfo{volume}{22}, \bibinfo{number}{5} (\bibinfo{year}{2015}), \bibinfo{pages}{69--71}.
\newblock


\bibitem[Lazar et~al\mbox{.}(2016)]%
        {lazar2016human}
\bibfield{author}{\bibinfo{person}{Jonathan Lazar}, \bibinfo{person}{Julio Abascal}, \bibinfo{person}{Simone Barbosa}, \bibinfo{person}{Jeremy Barksdale}, \bibinfo{person}{Batya Friedman}, \bibinfo{person}{Jens Grossklags}, \bibinfo{person}{Jan Gulliksen}, \bibinfo{person}{Jeff Johnson}, \bibinfo{person}{Tom McEwan}, \bibinfo{person}{Lo{\"\i}c Mart{\'\i}nez-Normand}, {et~al\mbox{.}}} \bibinfo{year}{2016}\natexlab{}.
\newblock \showarticletitle{Human--computer interaction and international public policymaking: a framework for understanding and taking future actions}.
\newblock \bibinfo{journal}{\emph{Foundations and Trends{\textregistered} in Human--Computer Interaction}} \bibinfo{volume}{9}, \bibinfo{number}{2} (\bibinfo{year}{2016}), \bibinfo{pages}{69--149}.
\newblock


\bibitem[Lee et~al\mbox{.}(2015)]%
        {lee2015working}
\bibfield{author}{\bibinfo{person}{Min~Kyung Lee}, \bibinfo{person}{Daniel Kusbit}, \bibinfo{person}{Evan Metsky}, {and} \bibinfo{person}{Laura Dabbish}.} \bibinfo{year}{2015}\natexlab{}.
\newblock \showarticletitle{Working with machines: The impact of algorithmic and data-driven management on human workers}. In \bibinfo{booktitle}{\emph{Proceedings of the 33rd annual ACM conference on human factors in computing systems}}. \bibinfo{pages}{1603--1612}.
\newblock


\bibitem[Licklider(1960)]%
        {licklider1960man}
\bibfield{author}{\bibinfo{person}{Joseph~CR Licklider}.} \bibinfo{year}{1960}\natexlab{}.
\newblock \showarticletitle{Man-computer symbiosis}.
\newblock \bibinfo{journal}{\emph{IRE transactions on human factors in electronics}} \bibinfo{number}{1} (\bibinfo{year}{1960}), \bibinfo{pages}{4--11}.
\newblock


\bibitem[Nagaraj~Rao et~al\mbox{.}(2025a)]%
        {rao2025quallm}
\bibfield{author}{\bibinfo{person}{Varun Nagaraj~Rao}, \bibinfo{person}{Eesha Agarwal}, \bibinfo{person}{Samantha Dalal}, \bibinfo{person}{Dan Calacci}, {and} \bibinfo{person}{Andr{\'e}s Monroy-Hern{\'a}ndez}.} \bibinfo{year}{2025}\natexlab{a}.
\newblock \showarticletitle{QuaLLM: An LLM-based framework to extract quantitative insights from online forums}.
\newblock \bibinfo{journal}{\emph{Findings of ACL, NAACL}} (\bibinfo{year}{2025}).
\newblock


\bibitem[Nagaraj~Rao et~al\mbox{.}(2025b)]%
        {rao2025rideshare}
\bibfield{author}{\bibinfo{person}{Varun Nagaraj~Rao}, \bibinfo{person}{Samantha Dalal}, \bibinfo{person}{Eesha Agarwal}, \bibinfo{person}{Dana Calacci}, {and} \bibinfo{person}{Andr{\'e}s Monroy-Hern{\'a}ndez}.} \bibinfo{year}{2025}\natexlab{b}.
\newblock \showarticletitle{Rideshare Transparency: Translating Gig Worker Insights on {AI} Platform Design to Policy}.
\newblock \bibinfo{journal}{\emph{Proceedings of the ACM on Human-Computer Interaction}} \bibinfo{volume}{9}, \bibinfo{number}{2} (\bibinfo{year}{2025}), \bibinfo{pages}{1--49}.
\newblock


\bibitem[Okamura et~al\mbox{.}(1995)]%
        {okamuraHelpingCSCWApplications1995}
\bibfield{author}{\bibinfo{person}{Kazuo Okamura}, \bibinfo{person}{Masayo Fujimoto}, \bibinfo{person}{Wanda~J. Orlikowski}, {and} \bibinfo{person}{Joanne Yates}.} \bibinfo{year}{1995}\natexlab{}.
\newblock \showarticletitle{Helping CSCW Applications Succeed: The Role of Mediators in the Context of Use}.
\newblock \bibinfo{journal}{\emph{The Information Society}} \bibinfo{volume}{11}, \bibinfo{number}{3} (\bibinfo{date}{July} \bibinfo{year}{1995}), \bibinfo{pages}{157--172}.
\newblock
\href{https://doi.org/10.1080/01972243.1995.9960190}{doi:\nolinkurl{10.1080/01972243.1995.9960190}}


\bibitem[Orlikowski and Baroudi(1991)]%
        {orlikowski1991studying}
\bibfield{author}{\bibinfo{person}{Wanda~J Orlikowski} {and} \bibinfo{person}{Jack~J Baroudi}.} \bibinfo{year}{1991}\natexlab{}.
\newblock \showarticletitle{Studying information technology in organizations: Research approaches and assumptions}.
\newblock \bibinfo{journal}{\emph{Information systems research}} \bibinfo{volume}{2}, \bibinfo{number}{1} (\bibinfo{year}{1991}), \bibinfo{pages}{1--28}.
\newblock


\bibitem[Rosenblat and Stark(2016)]%
        {rosenblatAlgorithmicLaborInformation2016b}
\bibfield{author}{\bibinfo{person}{Alex Rosenblat} {and} \bibinfo{person}{Luke Stark}.} \bibinfo{year}{2016}\natexlab{}.
\newblock \showarticletitle{Algorithmic Labor and Information Asymmetries: {{A}} Case Study of {{Uber}}’s Drivers}.
\newblock   \bibinfo{volume}{10} (\bibinfo{year}{2016}), \bibinfo{pages}{27}.
\newblock


\bibitem[Schwartz et~al\mbox{.}(2023)]%
        {schwartz2023deactivation}
\bibfield{author}{\bibinfo{person}{Lindsey Schwartz}, \bibinfo{person}{Nic Weber}, {and} \bibinfo{person}{Eva~Maxfield Brown}.} \bibinfo{year}{2023}\natexlab{}.
\newblock \showarticletitle{Deactivation with and without Representation: The Role of Dispute Arbitration for Seattle Rideshare Drivers}.
\newblock  (\bibinfo{year}{2023}).
\newblock


\bibitem[Service(2025)]%
        {IRS2025}
\bibfield{author}{\bibinfo{person}{Internal~Revenue Service}.} \bibinfo{year}{2025}\natexlab{}.
\newblock \bibinfo{title}{Independent contractor (self-employed) or employee?}
\newblock \bibinfo{howpublished}{Website}.
\newblock
\urldef\tempurl%
\url{https://www.irs.gov/businesses/small-businesses-self-employed/independent-contractor-self-employed-or-employee}
\showURL{%
\tempurl}
\newblock
\shownote{Accessed on 2025-05-03}.


\bibitem[Siddiqui(2021)]%
        {wapo2021deactivated}
\bibfield{author}{\bibinfo{person}{Faiz Siddiqui}.} \bibinfo{year}{2021}\natexlab{}.
\newblock \bibinfo{title}{Uber and Lyft to share data on driver deactivations, more than a year after pledging to do so}.
\newblock
\urldef\tempurl%
\url{https://www.washingtonpost.com/technology/2021/03/11/uber-lyft-driver-database/}
\showURL{%
\tempurl}


\bibitem[Spaa et~al\mbox{.}(2019)]%
        {spaa2019understanding}
\bibfield{author}{\bibinfo{person}{Anne Spaa}, \bibinfo{person}{Abigail Durrant}, \bibinfo{person}{Chris Elsden}, {and} \bibinfo{person}{John Vines}.} \bibinfo{year}{2019}\natexlab{}.
\newblock \showarticletitle{Understanding the Boundaries between Policymaking and HCI}. In \bibinfo{booktitle}{\emph{Proceedings of the 2019 CHI Conference on Human Factors in Computing Systems}}. \bibinfo{pages}{1--15}.
\newblock


\bibitem[Spaa et~al\mbox{.}(2022)]%
        {spaa2022creative}
\bibfield{author}{\bibinfo{person}{Anne Spaa}, \bibinfo{person}{Nick Spencer}, \bibinfo{person}{Abigail Durrant}, {and} \bibinfo{person}{John Vines}.} \bibinfo{year}{2022}\natexlab{}.
\newblock \showarticletitle{Creative and collaborative reflective thinking to support policy deliberation and decision making}.
\newblock \bibinfo{journal}{\emph{Evidence \& Policy}} \bibinfo{volume}{18}, \bibinfo{number}{2} (\bibinfo{year}{2022}), \bibinfo{pages}{376--390}.
\newblock


\bibitem[Star and Strauss(1999)]%
        {star1999layers}
\bibfield{author}{\bibinfo{person}{Susan~Leigh Star} {and} \bibinfo{person}{Anselm Strauss}.} \bibinfo{year}{1999}\natexlab{}.
\newblock \showarticletitle{Layers of silence, arenas of voice: The ecology of visible and invisible work}.
\newblock \bibinfo{journal}{\emph{Computer supported cooperative work (CSCW)}}  \bibinfo{volume}{8} (\bibinfo{year}{1999}), \bibinfo{pages}{9--30}.
\newblock


\bibitem[Suchman(1987)]%
        {suchman1987plans}
\bibfield{author}{\bibinfo{person}{Lucille~Alice Suchman}.} \bibinfo{year}{1987}\natexlab{}.
\newblock \bibinfo{booktitle}{\emph{Plans and situated actions: The problem of human-machine communication}}.
\newblock \bibinfo{publisher}{Cambridge university press}.
\newblock


\bibitem[Tang et~al\mbox{.}(2023)]%
        {tang2023back}
\bibfield{author}{\bibinfo{person}{Joice Tang}, \bibinfo{person}{McKane Andrus}, \bibinfo{person}{Samuel So}, \bibinfo{person}{Udayan Tandon}, \bibinfo{person}{Andr{\'e}s Monroy-Hern{\'a}ndez}, \bibinfo{person}{Vera Khovanskaya}, \bibinfo{person}{Sean~A Munson}, \bibinfo{person}{Mark Zachry}, {and} \bibinfo{person}{Sucheta Ghoshal}.} \bibinfo{year}{2023}\natexlab{}.
\newblock \showarticletitle{Back to “Back to Labor”: Revisiting Political Economies of Computer-Supported Cooperative Work}. In \bibinfo{booktitle}{\emph{Companion Publication of the 2023 Conference on Computer Supported Cooperative Work and Social Computing}}. \bibinfo{pages}{522--526}.
\newblock


\bibitem[Terry(2022)]%
        {cbs2022deactivated}
\bibfield{author}{\bibinfo{person}{Jermont Terry}.} \bibinfo{year}{2022}\natexlab{}.
\newblock \bibinfo{title}{Chicago Rideshare Drivers Say They're Being Deactivated Unfairly After Bad Reviews, Call For Hearings Before Deactivation}.
\newblock
\urldef\tempurl%
\url{https://www.cbsnews.com/chicago/news/chicago-rideshare-drivers-call-deactivation-hearings/}
\showURL{%
\tempurl}


\bibitem[Urquhart and Rodden(2017)]%
        {urquhart2017new}
\bibfield{author}{\bibinfo{person}{Lachlan Urquhart} {and} \bibinfo{person}{Tom Rodden}.} \bibinfo{year}{2017}\natexlab{}.
\newblock \showarticletitle{New directions in information technology law: learning from human--computer interaction}.
\newblock \bibinfo{journal}{\emph{International Review of Law, Computers \& Technology}} \bibinfo{volume}{31}, \bibinfo{number}{2} (\bibinfo{year}{2017}), \bibinfo{pages}{150--169}.
\newblock


\bibitem[Urquhart et~al\mbox{.}(2022)]%
        {urquhart2022legal}
\bibfield{author}{\bibinfo{person}{Lachlan~D Urquhart}, \bibinfo{person}{Glenn McGarry}, {and} \bibinfo{person}{Andy Crabtree}.} \bibinfo{year}{2022}\natexlab{}.
\newblock \showarticletitle{Legal provocations for HCI in the design and development of trustworthy autonomous systems}. In \bibinfo{booktitle}{\emph{Nordic Human-Computer Interaction Conference}}. \bibinfo{pages}{1--12}.
\newblock


\bibitem[Vasudevan and Chan(2022)]%
        {vasudevan2022gamification}
\bibfield{author}{\bibinfo{person}{Krishnan Vasudevan} {and} \bibinfo{person}{Ngai~Keung Chan}.} \bibinfo{year}{2022}\natexlab{}.
\newblock \showarticletitle{Gamification and work games: Examining consent and resistance among Uber drivers}.
\newblock \bibinfo{journal}{\emph{New Media \& Society}} \bibinfo{volume}{24}, \bibinfo{number}{4} (\bibinfo{year}{2022}), \bibinfo{pages}{866--886}.
\newblock


\bibitem[Watkins(2023)]%
        {watkinsFaceWorkHumanCentered2023}
\bibfield{author}{\bibinfo{person}{Elizabeth~Anne Watkins}.} \bibinfo{year}{2023}\natexlab{}.
\newblock \showarticletitle{Face Work: A Human-Centered Investigation into Facial Verification in Gig Work}.
\newblock \bibinfo{journal}{\emph{Proceedings of the ACM on Human-Computer Interaction}} (\bibinfo{date}{April} \bibinfo{year}{2023}), \bibinfo{pages}{1--24}.
\newblock
\href{https://doi.org/10.1145/3579485}{doi:\nolinkurl{10.1145/3579485}}


\bibitem[Whitney et~al\mbox{.}(2021)]%
        {whitney2021hci}
\bibfield{author}{\bibinfo{person}{Cedric~Deslandes Whitney}, \bibinfo{person}{Teresa Naval}, \bibinfo{person}{Elizabeth Quepons}, \bibinfo{person}{Simrandeep Singh}, \bibinfo{person}{Steven~R Rick}, {and} \bibinfo{person}{Lilly Irani}.} \bibinfo{year}{2021}\natexlab{}.
\newblock \showarticletitle{HCI tactics for politics from below: Meeting the challenges of smart cities}. In \bibinfo{booktitle}{\emph{Proceedings of the 2021 CHI conference on human factors in computing systems}}. \bibinfo{pages}{1--15}.
\newblock


\bibitem[Woodcock(2021)]%
        {woodcock2021towards}
\bibfield{author}{\bibinfo{person}{Jamie Woodcock}.} \bibinfo{year}{2021}\natexlab{}.
\newblock \showarticletitle{Towards a digital workerism: Workers’ inquiry, methods, and technologies}.
\newblock \bibinfo{journal}{\emph{NanoEthics}} \bibinfo{volume}{15}, \bibinfo{number}{1} (\bibinfo{year}{2021}), \bibinfo{pages}{87--98}.
\newblock


\bibitem[Woodcock and Johnson(2018)]%
        {woodcockGamificationWhatIt2018}
\bibfield{author}{\bibinfo{person}{Jamie Woodcock} {and} \bibinfo{person}{Mark~R. Johnson}.} \bibinfo{year}{2018}\natexlab{}.
\newblock \showarticletitle{Gamification: What It Is, and How to Fight It}.
\newblock \bibinfo{journal}{\emph{The Sociological Review}} \bibinfo{volume}{66}, \bibinfo{number}{3} (\bibinfo{date}{May} \bibinfo{year}{2018}), \bibinfo{pages}{542--558}.
\newblock
\href{https://doi.org/10.1177/0038026117728620}{doi:\nolinkurl{10.1177/0038026117728620}}


\bibitem[Yang et~al\mbox{.}(2024)]%
        {yang2024future}
\bibfield{author}{\bibinfo{person}{Qian Yang}, \bibinfo{person}{Richmond~Y. Wong}, \bibinfo{person}{Steven Jackson}, \bibinfo{person}{Sabine Junginger}, \bibinfo{person}{Margaret~D. Hagan}, \bibinfo{person}{Thomas Gilbert}, {and} \bibinfo{person}{John Zimmerman}.} \bibinfo{year}{2024}\natexlab{}.
\newblock \showarticletitle{The Future of HCI-Policy Collaboration}. In \bibinfo{booktitle}{\emph{Proceedings of the 2024 CHI Conference on Human Factors in Computing Systems}} (Honolulu, HI, USA) \emph{(\bibinfo{series}{CHI '24})}. \bibinfo{publisher}{Association for Computing Machinery}, \bibinfo{address}{New York, NY, USA}, Article \bibinfo{articleno}{820}, \bibinfo{numpages}{15}~pages.
\newblock
\showISBNx{9798400703300}
\href{https://doi.org/10.1145/3613904.3642771}{doi:\nolinkurl{10.1145/3613904.3642771}}


\bibitem[Zhang et~al\mbox{.}(2023)]%
        {zhang2023stakeholder}
\bibfield{author}{\bibinfo{person}{Angie Zhang}, \bibinfo{person}{Alexander Boltz}, \bibinfo{person}{Jonathan Lynn}, \bibinfo{person}{Chun-Wei Wang}, {and} \bibinfo{person}{Min~Kyung Lee}.} \bibinfo{year}{2023}\natexlab{}.
\newblock \showarticletitle{Stakeholder-Centered AI Design: Co-Designing Worker Tools with Gig Workers through Data Probes}. In \bibinfo{booktitle}{\emph{Proceedings of the 2023 CHI Conference on Human Factors in Computing Systems}}. \bibinfo{pages}{1--19}.
\newblock


\end{thebibliography}

\appendix

\section{Rideshare Drivers' Experiences with Deactivations and Reactivations}
 There is limited prior research on rideshare driver deactivations. To provide context and ground the study, we draw on the only existing empirical work on this topic in Seattle, WA \cite{schwartz2023deactivation}, which analyzed a sample of 1,420 deactivated drivers. To enhance this localized focus, we also draw on broader insights from Reddit discussions (r/uberdrivers and r/lyftdrivers), which provide a complementary perspective on rideshare drivers’ experiences, and help contextualize local findings within shared challenges faced by a diverse group of drivers. We leverage the QuaLLM framework \cite{rao2025quallm, rao2025rideshare} and the LLM-based tool \url{https://thegigabrain.com/} for this analysis. Drawing on these resources, we now summarize drivers' experiences with deactivation causes, challenges in reactivation, strategies for reactivation, and preventive measures.
 
 \subsection{Drivers Are Deactivated by Arbitrary AI, Algorithms, and Occasionally Human Review:}
Uber and Lyft rely on automated surveillance systems, third-party hiring platforms (e.g. HireRight and Checkr) and passenger feedback to algorithmically manage rideshare drivers. When these systems flag issues, drivers can be suddenly deactivated by an in-app pop-up, SMS, or email \cite{schwartz2023deactivation} as seen in Figure \ref{fig:reddit-deactivations}. While these platforms have implemented some deactivation processes, they lack standardized classifications and clear thresholds for violations that result in driver deactivations. 

Despite this inconsistency, three common causes of deactivations have emerged across these platforms: passenger and vehicle safety, background checks and documentation, and account verification. Passenger safety concerns often involve complaints such as harassment, driving under the influence, speeding, or refusing service to passengers with service animals. Background check issues typically relate to problems with licenses or permits, such as pending criminal charges. Account verification issues include allegations of fraudulent activity, such as duplicate accounts. Historically, drivers received minimal notice of deactivation and were denied due process until an ordinance was implemented in 2021 and subsequent laws (described below) provided protections.
 
\subsection{Reactivation is Often a Long and Futile Process:}
When deactivated due to passenger complaints, drivers are often not provided even basic anonymized details about the trip in question. This lack of transparency prevents them from defending themselves against false allegations. Many drivers report that the platform’s support channels—whether via messages, email, or in-person visits—are unresponsive or dismissive. Detailed appeals are frequently ignored, with the platform adhering to its initial decisions. Some drivers have found that publicly posting on the platform’s social media pages can expedite resolution by pressuring the company to act. Others have pursued legal remedies through arbitration or small claims court to contest their deactivations, which could create financial hurdles for drivers.

\subsection{Drivers Employ Dashcams, Follow Guidelines, and Document Incidents to Prevent Deactivation}
To mitigate the risk of deactivation, drivers adopt various preventive measures. Using dashcams to record rides provides evidence in case of false passenger reports. Familiarizing themselves with Uber’s community guidelines and adhering strictly to them helps avoid violations. Clear communication during rides is another key strategy; documenting incidents immediately can preempt passenger complaints and provide a record of events if disputes arise.

\section{Backend Architecture}
The backend processes handle data collection, storage, and processing efficiently. FairFare is an open-source web application built with Svelte\footnote{Svelte is a ``free and open-source component-based front-end software framework'' available at \url{https://svelte.dev/}} and hosted on Netlify. The system uses Twilio to manage OTP delivery to users' phone numbers and Supabase as the database backend.

Argyle data syncs to the database through webhooks with daily refreshes. According to Argyle's documentation, ``when a payroll account [e.g. Uber] is connected..., all available data and documents are retrieved and processed. The information is digitized, standardized, and made accessible..., alongside the original downloadable raw document files''\footnote{\url{https://argyle.com/docs/overview/data-structure/data-sets}}. This integration provides comprehensive access to drivers' historical earnings data without requiring manual screenshot collection.

\section{Formative Study Interview Protocol}
\label{app:formative-study-protocol}
\subsection*{Introduction (10 minutes)}
Today's activity will be divided into four main parts, with approximately 5--10 minutes allocated per part:
\begin{itemize}
    \item Policy Context
    \item Current Process and Implementation
    \item Defining Success
    \item Challenges
\end{itemize}

Broadly, the questions of interest are:
\begin{itemize}
    \item What is the policy context which enables \LU{} to file claims on behalf of deactivated drivers?
    \item What is the typical process you follow currently to make claims about lost wage?
    \item How do you measure success in a deactivation case?
    \item What are some of the challenging aspects of this?
\end{itemize}

\subsection*{Part 1: Policy Context (10 minutes)}
\textbf{Transition:} Now that we've introduced ourselves, let's start by understanding the broader policy landscape that shapes your work. We're particularly interested in learning about how Seattle's regulations have influenced your approach to handling driver deactivations.

\begin{enumerate}
    \item How has the Seattle policy shaped your approach to handling deactivation cases?
    \item What requirements or obligations did this policy create for your organization?
    \item How did \LU{} become the relevant point of contact to support drivers? How were these requirements communicated to you?
\end{enumerate}

\subsection*{Part 2: Current Process and Implementation (10 minutes)}
\textbf{Transition:} Thank you for helping us understand the policy framework. Now we'd like to dive into the day-to-day aspects of your deactivation work.

\begin{enumerate}
    \item Could you walk us through the typical process when a driver approaches you about a deactivation?
    \item How do you currently document and calculate back-pay claims?
    \item What resources or tools does your team use to manage deactivation cases?
    \item What aspects of the process tend to be most time-consuming or resource-intensive?
    \item How did you develop your current process for handling these cases?
\end{enumerate}

\subsection*{Part 3: Defining Success (10 minutes)}
\textbf{Transition:} We've gotten a good picture of your current process. Let's shift our focus to understanding what success looks like in these cases and how you measure it.

\begin{enumerate}
    \item How does your organization define success in a deactivation case? What specific metrics or markers do you use to measure success?
    \item Could you describe a recent case that you would consider particularly successful?
    \item Why do you consider it successful? What factors contributed to that success?
\end{enumerate}

\subsection*{Part 4: Challenges (10 minutes)}
\textbf{Transition:} Having discussed what success looks like, we'd like to understand the obstacles you face in achieving these goals. Let's talk about some of the challenges in your current process.

\begin{enumerate}
    \item What aspects of the current deactivation process do you find most challenging and would like to improve?
    \item What makes that particularly challenging?
\end{enumerate}

\subsection*{Conclusions (5 minutes)}
We're nearing the end of our time\dots

\begin{enumerate}
    \item Is there anything else about the deactivation claims process that you think is important for us to understand?
    \item Do you have any other questions for us?
\end{enumerate}

Thank you for sharing your experiences with us. Your insights will help us better understand how organizations like yours implement these policies and support drivers through the deactivation process.

\section{Field Visit Schedule}
\label{app:field-visit-schedule}
\begin{itemize}
    \item \textbf{Morning Session (9–11 AM):}
    \begin{itemize}
        \item Observed 7 field representatives conducting intake interviews with 10 drivers.
        \item Observations informed semi-structured interview questions about workflow challenges.
    \end{itemize}
    
    \item \textbf{Midday Session (11 AM–12 Noon):}
    \begin{itemize}
        \item Facilitated a focus group with 7 field representatives and 1 lead organizer to further explore their experiences with the tool.
    \end{itemize}
    
    \item \textbf{Afternoon Session (1–1:30 PM):}
    \begin{itemize}
        \item Observed 2 attorneys using the tool to generate lost wage reports.
        \item Administered a system usability survey to the 2 attorneys to gather feedback on the tool’s usability.
    \end{itemize}
    
    \item \textbf{Late Afternoon Session (1:30–4 PM):}
    \begin{itemize}
        \item Held a focus group with 2 attorneys and 1 lead organizer to collect qualitative feedback on how the tool impacted their workflows.
    \end{itemize}
    \item \textbf{Evening Session (4-5 PM):}
    \begin{itemize}
        \item Held a technical debugging session with 2 attorneys and 1 lead organizer to resolve any issues, improve communication between DU staff and our research team, and establish escalation processes for ongoing support.
    \end{itemize}
\end{itemize}

\section{Field Visit: Study Protocol for \TOOL{} Evaluation}
\label{app:field-visit-protocol}

\subsection*{Pre-Study Questions (5 minutes)}

\textbf{Background:}
\begin{enumerate}
    \item How long have you worked with \LU{}?
    \item What is your experience level with the tool?
    \item How long have you been using the automated lost wage calculation tool?
    \item On average, how many cases do you handle per week?
    \item Self-reported time for report generation pre-tool vs. with tool:
    \begin{itemize}
        \item Before using the tool, approximately how long did it take you to calculate the unemployment compensation?
        \item Since using the tool, approximately how long does it take you to calculate the unemployment compensation? (You can exclude data syncing time.)
    \end{itemize}
\end{enumerate}

\subsection*{Observation Task}

\textit{Aim: Observe at least 5 each of intake and report generation processes.}

\subsubsection*{Task 1: Intake Process}

The researcher will observe the organizer as they walk a driver through the intake process using FairFare, including:
\begin{itemize}
    \item Explaining the tool to the driver.
    \item Guiding the driver through signing up for FairFare.
    \item Ensuring all required information is entered into FairFare (until they receive the text message).
\end{itemize}

\textbf{Data Collected:}
\begin{itemize}
    \item Time taken to complete intake.
    \item Number of errors or missing fields during data entry.
    \item Verbal comments or feedback from the organizer during the task.
\end{itemize}

\textbf{Semi-structured Interview Questions (immediately after intake observation):}
\begin{enumerate}
    \item How does this compare to your previous intake workflow/data gathering? 
    \item How has FairFare impacted your ability to gather driver information?
    \item Has FairFare reduced the need for follow-ups with drivers for missing information?
    \item What challenges did you face during this process?
\end{enumerate}

\subsubsection*{Task 2: Report Generation Process}

The researcher will observe the organizer generate a lost wage report based on pre-synced data on \TOOL{}. The task will include:
\begin{itemize}
    \item Logging into the \TOOL{} Organizer Dashboard.
    \item Navigating to a driver's case file whose data has already synced.
    \item Inputting key details for report generation:
    \begin{itemize}
        \item Deactivation period
        \item Reference period
        \item Platform (e.g., Uber, Lyft)
    \end{itemize}
    \item Generating and downloading both PDF and CSV reports.
    \item If a driver is present, observe whether they are able to look through and manually verify synced trip data / are confident that data is correct.
\end{itemize}

\textbf{Data Collected:}
\begin{itemize}
    \item Time taken to generate a report.
    \item Number of errors or corrections needed during data entry.
    \item Verbal comments or feedback from participants during the task.
\end{itemize}

\textbf{Follow-Up Questions (immediately after report generation observation):}
\begin{enumerate}
    \item How does this compare to your previous method for calculating lost wages?
    \item Have you encountered situations where trip data was incomplete or incorrect? How did you handle it?
    \item Were there any parts of this process that were unclear or cumbersome?
\end{enumerate}

\subsection*{Usability Scale and Feedback (15 minutes)}

\begin{figure}[htb!]
    \centering
    \includegraphics[width=0.6\linewidth]{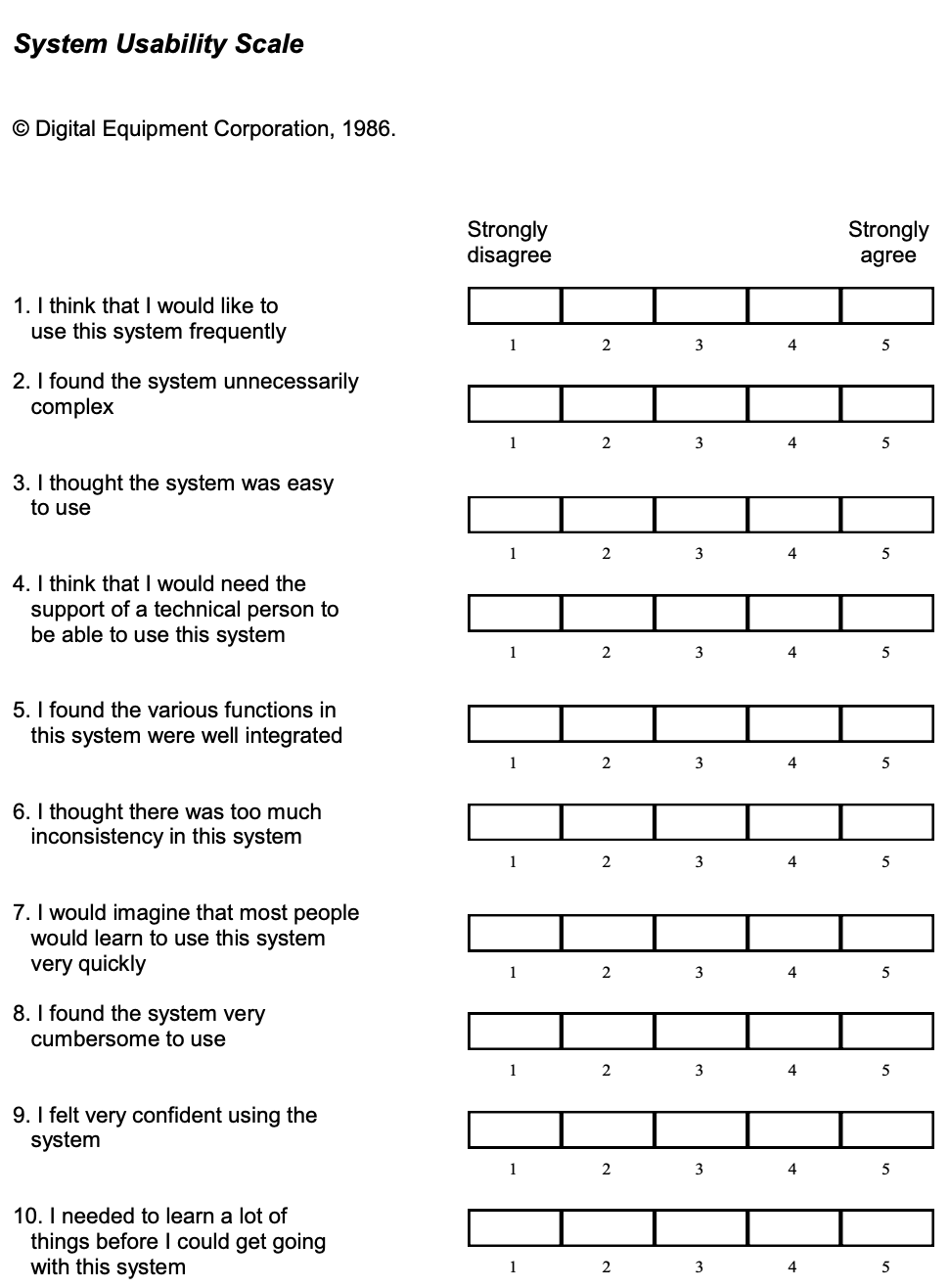}
    \caption{System Usability Scale from \cite{brooke1996sus}}
    \label{fig:sus}
\end{figure}

Attorneys will complete a System Usability Scale (SUS) questionnaire (see Figure \ref{fig:sus}) and share any other feedback. Then, ask the following questions:
\begin{enumerate}
    \item How has/how do you think \TOOL{} impacted/can impact the overall timeline for resolving cases?
    \item How has the tool influenced your communication with drivers about their cases?
    \item How confident are you in the accuracy of synced trip data?
    \item How has \TOOL{} impacted your ability to manage a high volume of cases?
    \item What aspect of the flow can be improved? (Ask about UI aspects)
\end{enumerate}

\end{document}